\documentclass[iop,apj]{emulateapj_may2010}
\usepackage{graphicx}
\usepackage{rotating} 

\def\gtrsim{\lower2pt\hbox{$\buildrel {\scriptstyle >}
   \over {\scriptstyle\sim}$}}
\def\lesssim{\lower2pt\hbox{$\buildrel {\scriptstyle <}
   \over {\scriptstyle\sim}$}}

\def\asca{{\sl ASCA }}

\def\xte{{\sl RXTE }}

\def\chandra{{\sl Chandra }}

\def\cirx1{Cir~X-1~}
\def\b0614{4U~0614+091~}
\def\u1626{4U~1626-67~}
\def\ergsec{\hbox{erg s$^{-1}$ }}

\def\kms{\hbox{km s$^{-1}$}}

\def\ha{H~{$\alpha$}}

\def\hei{He~{\sc i}}
\def\heii{He~{\sc ii}}
\def\nex{Ne~{\sc x}}
\def\neix{Ne~{\sc ix}}
\def\mgxi{Mg~{\sc xi}}
\def\mgxii{Mg~{\sc xii}}
\def\fexxvi{Fe~{\sc xxvi}}
\def\oviii{O~{\sc viii}}
\def\ovii{O~{\sc vii}}
\def\sixiv{Si~{\sc xiv}}
\def\Msun{$M_{\odot}$}
\def\Rsun{$R_{\odot}$ }

\def\kms{\hbox{km $\rm s^{-1}$}}

\slugcomment{Submitted to the Astrophysical Journal}

\shorttitle{X-Ray Lines in 4U1822$-$371}
\shortauthors{Ji et al.}

\begin{document}

\title{Implications of X-Ray Line Variations for 4U1822$-$371}

\author{L. Ji\altaffilmark{1}, N. S. Schulz\altaffilmark{1},\
        M. A. Nowak\altaffilmark{1}, C. R. Canizares\altaffilmark{1},
        }

\altaffiltext{1}{MIT Kavli Institute for Astrophysics \& Space Research}

\begin{abstract}
4U 1822$-$371 is one of the proto-type accretion disk coronal sources
with an orbital period of about 5.6 hours. The binary is viewed almost
edge-on at a high inclination angle of ~83 degrees, which makes it a
unique candidate to study binary orbital and accretion disk dynamics
in high powered X-ray sources. We observed the X-ray source in 4U
1822$-$371 with the Chandra High Energy Transmission Grating
Spectrometer (HETGS) for almost nine binary orbits. X-ray eclipse
times provide an update of the orbital ephemeris. We find that our
result follows the quadratic function implied by previous
observations; however, it suggests a flatter trend. Detailed line
dynamics also confirm a previous suggestion that the observed
photo-ionized line emission originates from a confined region in the
outer edge of the accretion disk near the hot spot. Line properties
allow us to impose limits on the size of accretion disk, the central
corona, and the emission region. The photo-ionized plasma is
consistent with ionization parameters of log $\xi > $ 2, and when
combined with disk size and reasonable assumptions for the plasma
density, this suggests illuminating disk luminosities which are over
an order of magnitude higher than what is actually observed.  That is,
we do not directly observe the central emitting X-ray source. The
spectral continua are best fit by a flat power law with a high energy
cut-off and partial covering absorption ($N_{\rm H}$ ranging from
5.4--$6.3\times 10^{22}\,{\rm cm}^{-2}$) with a covering fraction of
about 50$\%$. We discuss some implications of our findings with
respect to the photo-ionized line emission for the basic properties of
the X-ray source.
\end{abstract}


\section{INTRODUCTION}

Eclipsing X-ray binaries are still fairly rare as they require a view
of the accretion disk that is within several degrees of the edge. The
low-mass X-ray binary (LMXB) 4U 1822$-$371 is a specifically rare
object because it also has a relatively short orbital period which
produces eclipses every 5.57 h and a light curve which is very
sensitive to the structure of the accretion disk
rim~\citep{white1982}.  Repetitive features in the light curve as well
as the fact that the eclipse is only partial provide some direct
geometrical constraints such as a viewing angle $i$ within
$75^{\circ}$ and $85^{\circ}$ as well as a disk radius of
$(5-7.3)\times10^{10}$ cm for this range of viewing angles
\citep{mason1982, hellier1989}. \citet{heinz2001} determined the most
accurate inclination of $i = 82.5\pm1.5$ from {\it Rossi X-Ray Timing
  Explorer (RXTE)} data.

Its generally hard X-ray spectrum is an indicator that there is an
accretion disk corona (ADC) in the line of sight to the central X-ray
source which also partially obscures the source with a radial extent
of $\sim3\times10^{10}$ cm~\citep{white1981, white1982}.  X-ray
spectra from the pre-\chandra era were complex to model as they seem
to require additional and unresolved soft emission line structures in
addition to a significant Fe K fluorescence line. This made 4U
1822$-$371 the archetypical example of a LMXB ADC
source~\citep{white1981, white1997, parmar2000}.  The first \chandra
observation in 2000 finally resolved these line structures into
discrete emission lines from photo-ionized O, Ne, Mg, Si, S, and Fe
ions~\citep{cottam2001}.  Phase-resolved spectra also suggested that
the line emission originates from a highly confined area, likely the
illuminated hot spot. From the persistence of the Fe K fluorescence
line throughout the binary orbit, \citet{cottam2001} also concluded
that its emission must come from a more extended region around the
disk. A recent survey of all X-ray binaries in the \chandra archive
indeed shows that in nearly all cases narrow line fluorescence
originates from more or less spherically distributed material around
the centrally illuminating source~\citep{torrejon2010}.

More detailed dynamical information in LMXBs are hard to come by
simply because a clear identification of the low-mass companion at
distances of several kpc is usually impossible. \citet{jonker2001}
detected 0.59 s pulsations from the neutron star in 4U 1822$-$371 and
determined a donor star mass of 0.4 \Msun~ assuming a neutron star
mass of 1.4~\Msun.  From pulsation time delays they also determined
that the orbit must be circular. By observing Bowen fluorescence from
the X-ray heated face of the companion star, \citet{casares2003}
determined a radial velocity semi-amplitude of 300 \kms, which yielded
a companion mass of 0.36 \Msun.  However~\citet{cowley2003} used
strong \hei~ absorption to measure a lower limit of the orbital
velocity of the donor star to 234 \kms, consistent with previous
measurements~\citep{harlaftis1997}, providing a lower limit to the
donor mass of again 0.4 \Msun. In addition, \citet{cowley2003} also
measured \heii~ and \ha~ emission line velocities and determined a
systemic velocity of -103 \kms. They argued that such a large negative
value in direction of the Galactic center might indicate that 4U
1822$-$371 is either part of the halo population or it was given a
significant kick when the neutron star was formed.

In this paper we use phase- and highly spectrally resolved X-ray
spectra over almost nine binary orbits in order to further investigate
the origin of the recombination line emission and use their dynamics
to diagnose properties of the accretion disk.

\section{Data Reduction}

We observed the source 4U 1822$-$371 on 2008 May 20 (ID9076) and 23
(ID9858) with the \chandra HETGS \citep{canizares00} in a standard
configuration via the HETG Guaranteed Time Observation program (see
table 1).  With a total exposure time approximately 150\,ksec, these
two observations cover seven (three and four, respectively) full
binary periods.  We reprocess all the observations using CIAO Version
4.2 with the most recent CALDB products.

For each observation, we redetermine the positions of the zeroth-order
centroid using the {\tt findzo} script for optimal wavelength
accuracy.  We extract all first-order spectra in HEG and MEG and
co-add them into one single spectrum. This has the consequence that we
adopt a spectral resolution near 0.021\,m\AA\ provided by the medium
energy transmission grating grid throughout the entire wavelength
band. The sacrificed resolution is a trade-off for easier handling of
the spectra with respect to signal-to-noise ratios, specifically for
the phase-resolved spectral analysis.  Light curves and spectral
analysis are based on ISIS Versions 1.5.0--1.6.1 \citep{houck00}.

\section{Analysis and Results}

Eclipses are rare in LMXBs. They, however, contain critical dynamical
information with respect to the geometry of the disk and its
environment. A broad eclipse, for example, would indicate an extended
companion. A partial eclipse may indicate the presence of large scale
scattering structures such as ADC.  In a first step in the analysis we
derive an improved ephemeris based on all three HETGS observations and
test if the orbital period is changing. A second step investigates
orbital phase variations of bright X-ray lines. In a third step we fit
a phase-binned spectrum at an orbital phase which allows a direct view
of the photo-ionized region with appropriate continuum and
photo-ionization models.

\subsection{An Ephemeris Update}
\begin{figure}
\mbox{
\includegraphics[width=3.6in]{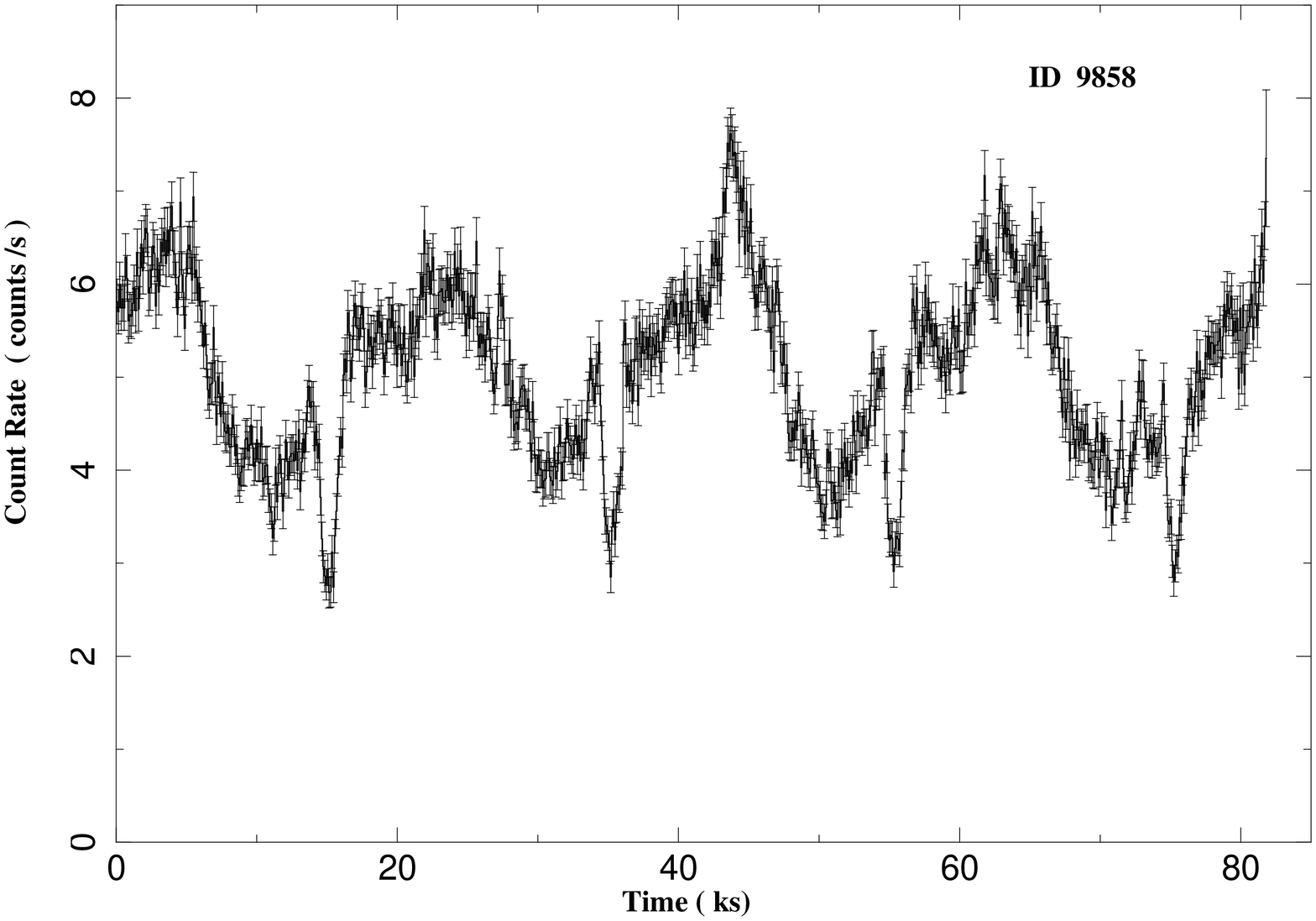}
\includegraphics[width=2.8in]{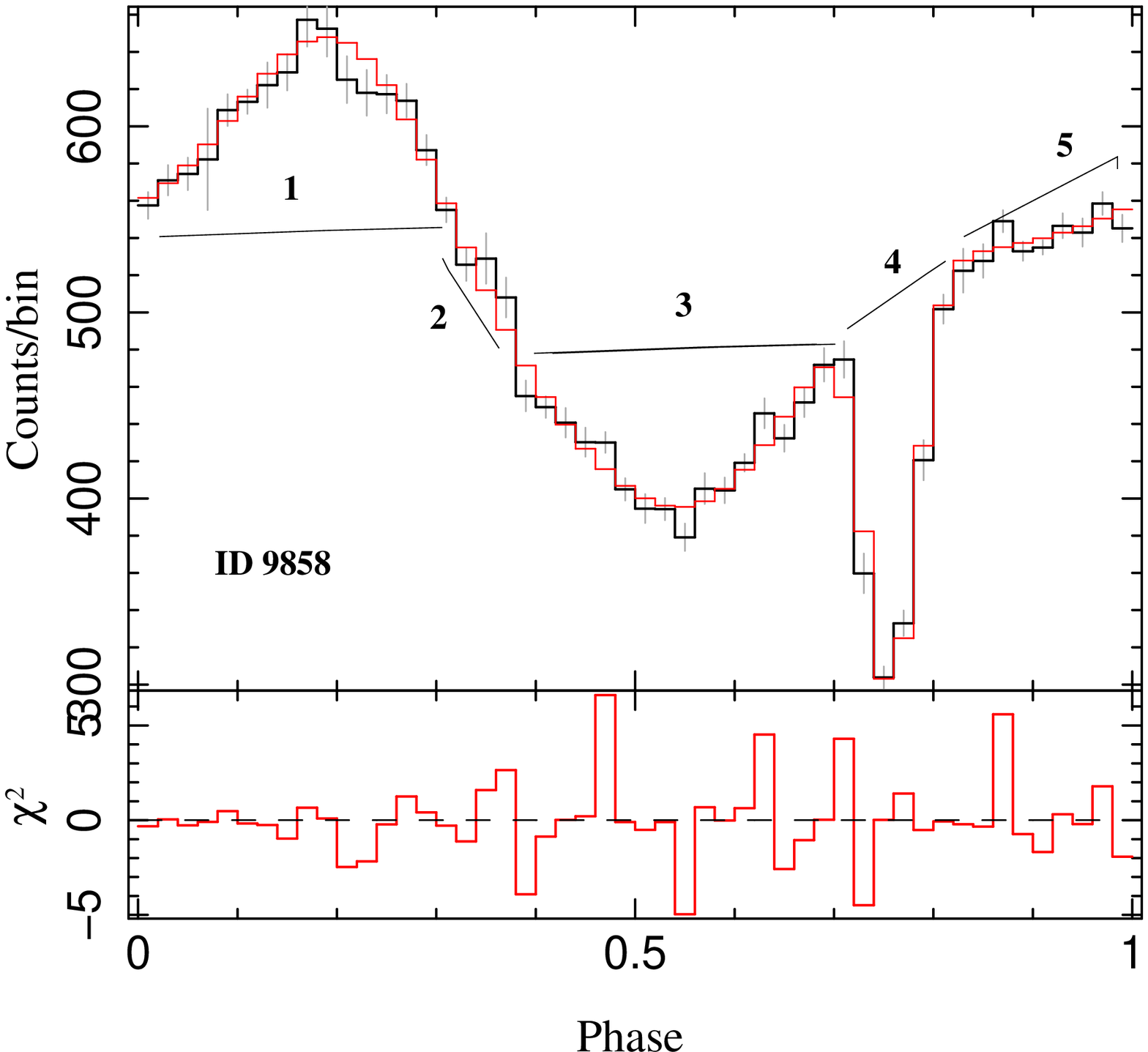}
}
\mbox{
\includegraphics[width=3.8in]{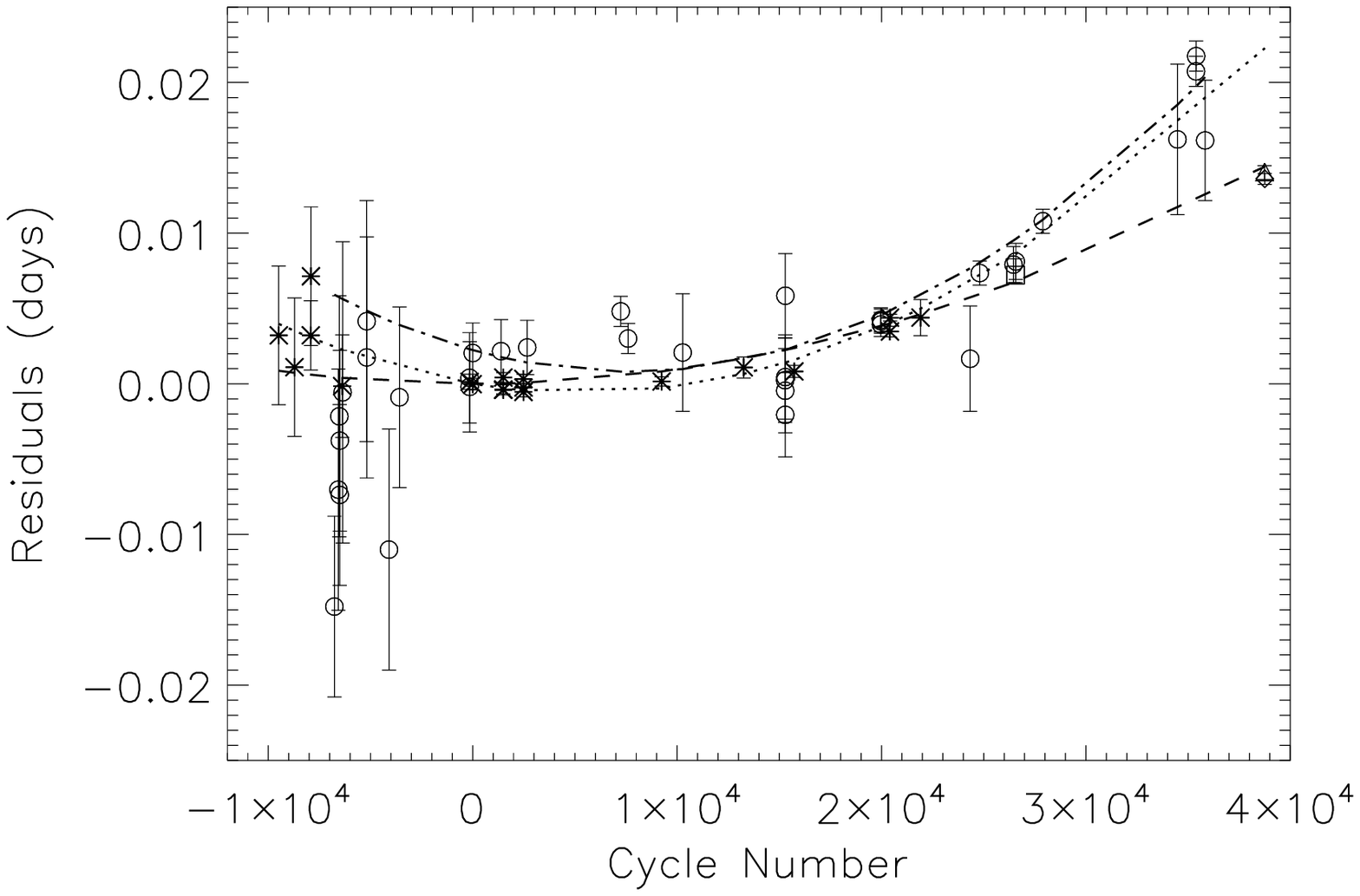}
\hspace*{-0.2in}
\includegraphics[width=2.7in]{newobs_pfold_new.ps}
}
\caption{ {\bf Upper left} --- Light curve for ID 9858 at wavelengths
  $\lambda > 1.5 ~\AA $; {\bf Upper right} --- The fitted folded phase
  light curve of ID9858: black --- observational curve; red --- fitted
  curve.  Short black lines and index mark the phase bins for Fe lines
  analysis.  {\bf Lower left}--- The 4U1822$-$371 partial eclipse
  timing residuals with respect to the best-fit linear ephemeris:
  previous X-ray eclipse measurements are at 68\% confidence levels
  --- asterisk \citep[][and reference therein]{parmar2000}; previous
  UV/Optical eclipse measurements at 90\% confidence --- circle; new
  measurements at 90\% confidence( square --- ID 671; triangle ---
  ID9076; diamond --- ID9858).  The best quadratic ephemeris from our
  measurement (dash line) is indicated: dotted line ---
  \citet{parmar2000}; dash-dotted line --- UV/Optical fitting.  {\bf
    Lower right}--- Phase folded light curves of ID9858 (black) \&
  ID9076 (red) with example out of eclipse phase bin width shown by
  blue shadows, and the in eclipse phase bin shown by the orange
  shadows.}
\label{fig-ephemeris}
\end{figure}

All existing HETGS observations of 4U 1822$-$371 yield a total of 9
complete eclipse light curves. The year 2000 data indicated a slowly
increasing orbital period.  Here we first folded each of the observations
into phase bins using a fixed period as determined by
\citet{parmar2000}. Then we fitted each phase-folded light curve with
Gaussian and sinusoidal functions. The results are shown in the right
upper panel of Figure \ref{fig-ephemeris} for ID9858 as an
illustration. The new eclipse time can be well-determined. Once we
join this new eclipse time to the already published X-ray eclipse
times \citep[][and the reference therein]{parmar2000}, we can fit the
timing residuals with respect to the best-fit linear ephemeris
($\chi^{2}=35.99 $ for 19 dof) using the quadratic function (shown in
left lower panel of Figure \ref{fig-ephemeris}). The new curve is
somewhat flatter than the old one and leads to an updated quadratic
ephemeris given by:
\begin{eqnarray}
 T_{\rm elc}= 2445615.30927(41) + 0.232109006(67) N \\ 
   \nonumber -   0.96(18)\times 10^{-11} N^{2}
\end{eqnarray}
Where N is the cycle number and linear ephemeris, and errors are at
90\% confidence on the last 2 digits.  For comparison, the dotted line
shows the quadratic fit obtained for the old X-ray datasets by
\citet{parmar2000}.  The new X-ray eclipse times derived from the HETG
observations are given in table \ref{tab-eclipse} with uncertainties
given at 90\% confidence.

\subsection{Phase Resolved Line Properties}

Observed spectral features are highly phase-dependent. The new
observations cover almost four times the number of binary orbits than
what \citet{cottam2001} had available, which allow us to investigate
line variations with orbital phase in much more detail. The lower
right panel in Figure \ref{fig-ephemeris} shows the phase separation
for our stronger line detections.  We center the first phase bin at
eclipse minimum.  To study the H-like and He-like lines, except the
narrow phase bin ($\sim 0.08$) during eclipse, we choose a phase bin
width of 0.16. This secures enough counts for the line features in
each bin, and the \nex, \neix, \mgxii, \mgxi, \oviii, and \ovii
~emission lines are clearly detected with this choice.  We divide the
orbit into 40 even, but overlapping, phase segments.  The \fexxvi~ and
Fe K$\alpha$ have very weak signal with this choice of (correlated)
phase bin, consequently, we additionally chose a set of five,
non-overlapping phase bins as marked in upper right panel of Figure
\ref{fig-ephemeris}.  They are denoted as upper(1), decreasing(2),
bottom(3), eclipsing(4), and rising(5) respectively.

\subsubsection{Fe lines}

The crude five phase binning of the \fexxvi~ and Fe K$\alpha$ ~lines
smears out most orbital information. Detections are shown in Table
\ref{tab-fe}, and no line shifts are found.  However, their flux
trends follow the shape of the light curves, reaching a minimum during
eclipse and a maximum at the light curve's maximum.  For all five
phases, the centroid of the Fe K line is consistent with
1.93~\AA\ which covers wavelengths of Fe~I to Fe~XX from the cold,
near-neutral medium, similar to the finding by~\citet{cottam2001}.
However, we find a lower limit to the line broadening of about $370~
\rm km~s^{-1}$.

\subsubsection{Bright Line Dynamics}

\begin{figure}
\mbox{
\includegraphics[width=2.4in]{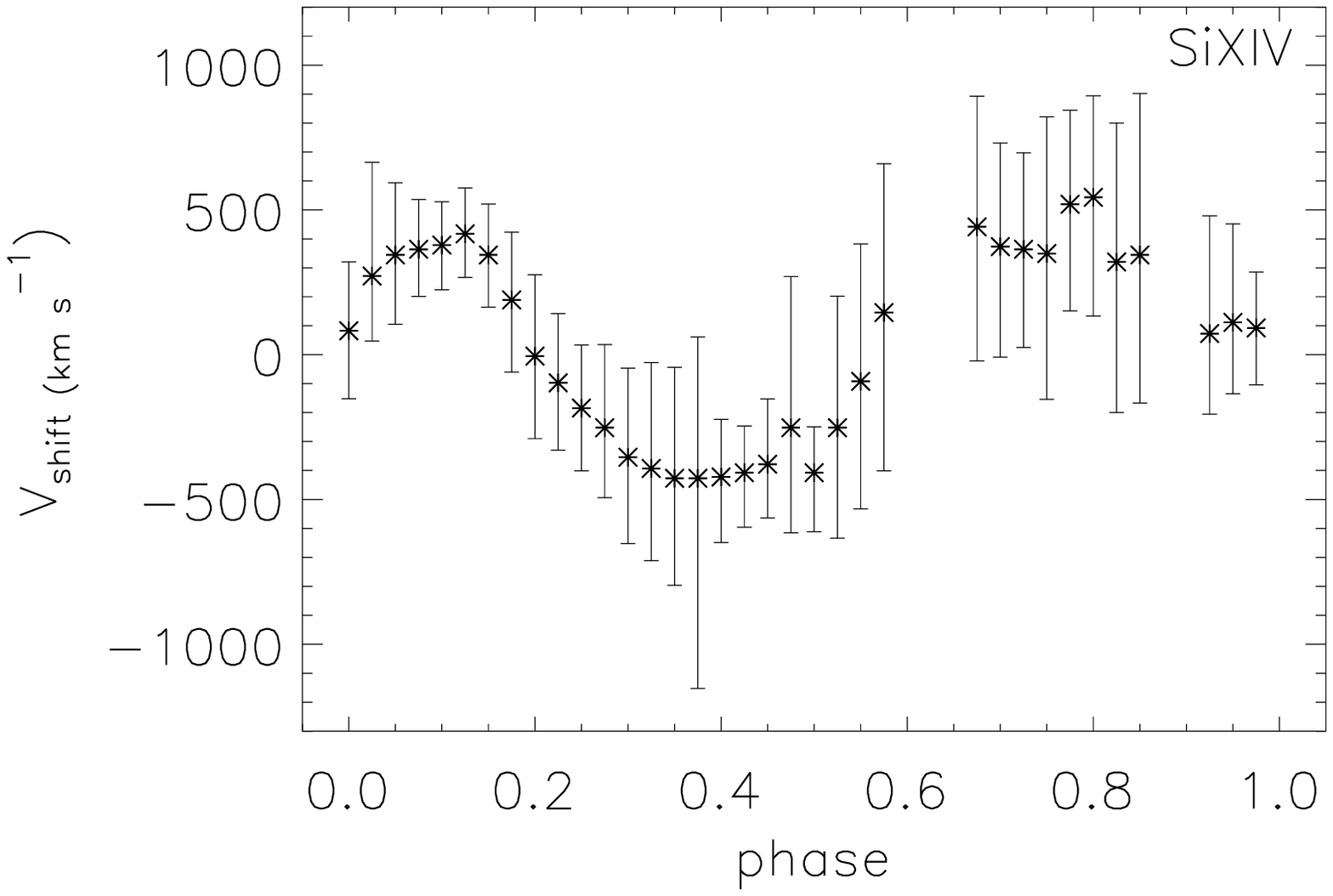}
\hskip -0.28in
\includegraphics[width=2.4in]{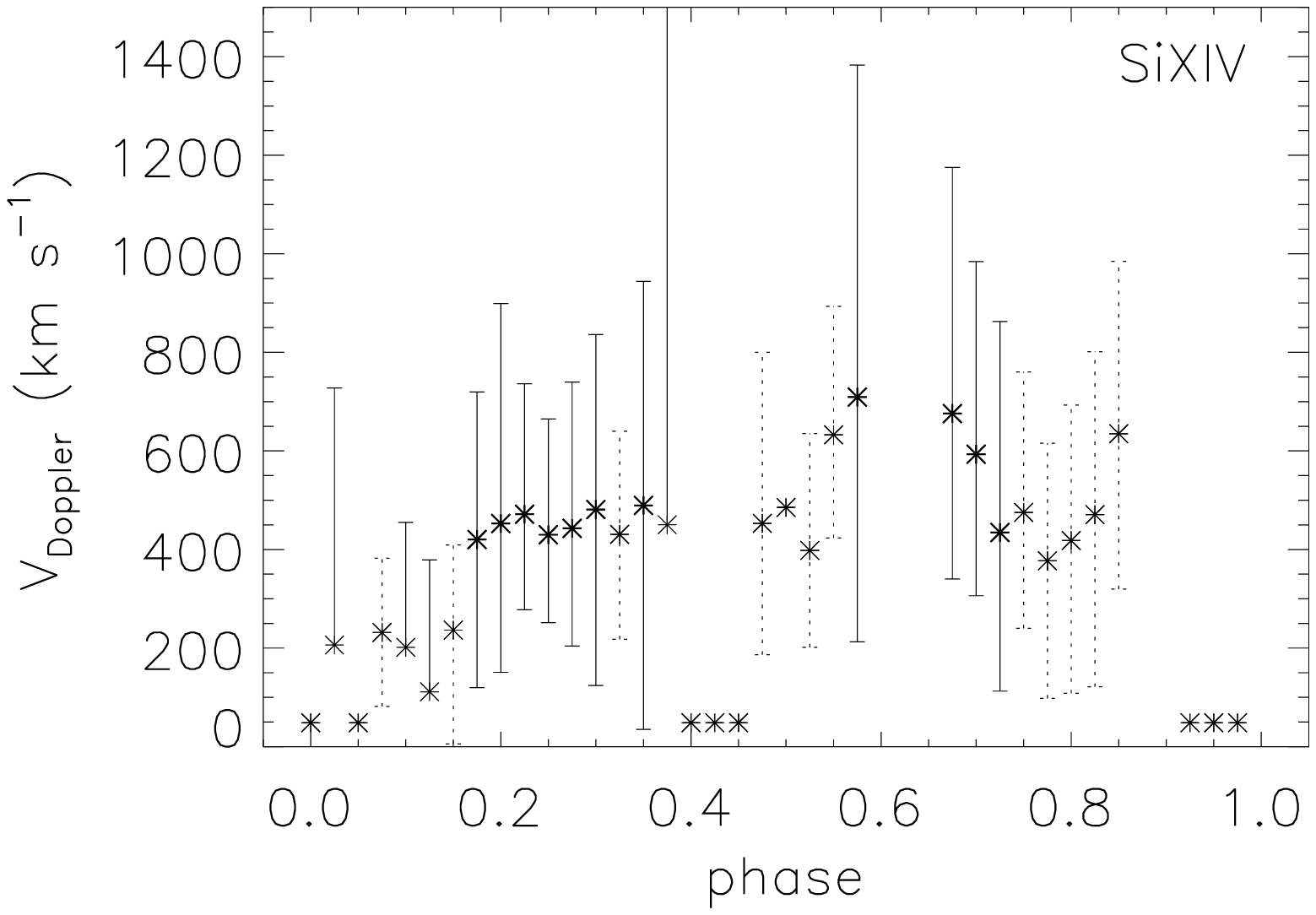}
\hskip -0.3in
\includegraphics[width=2.4in]{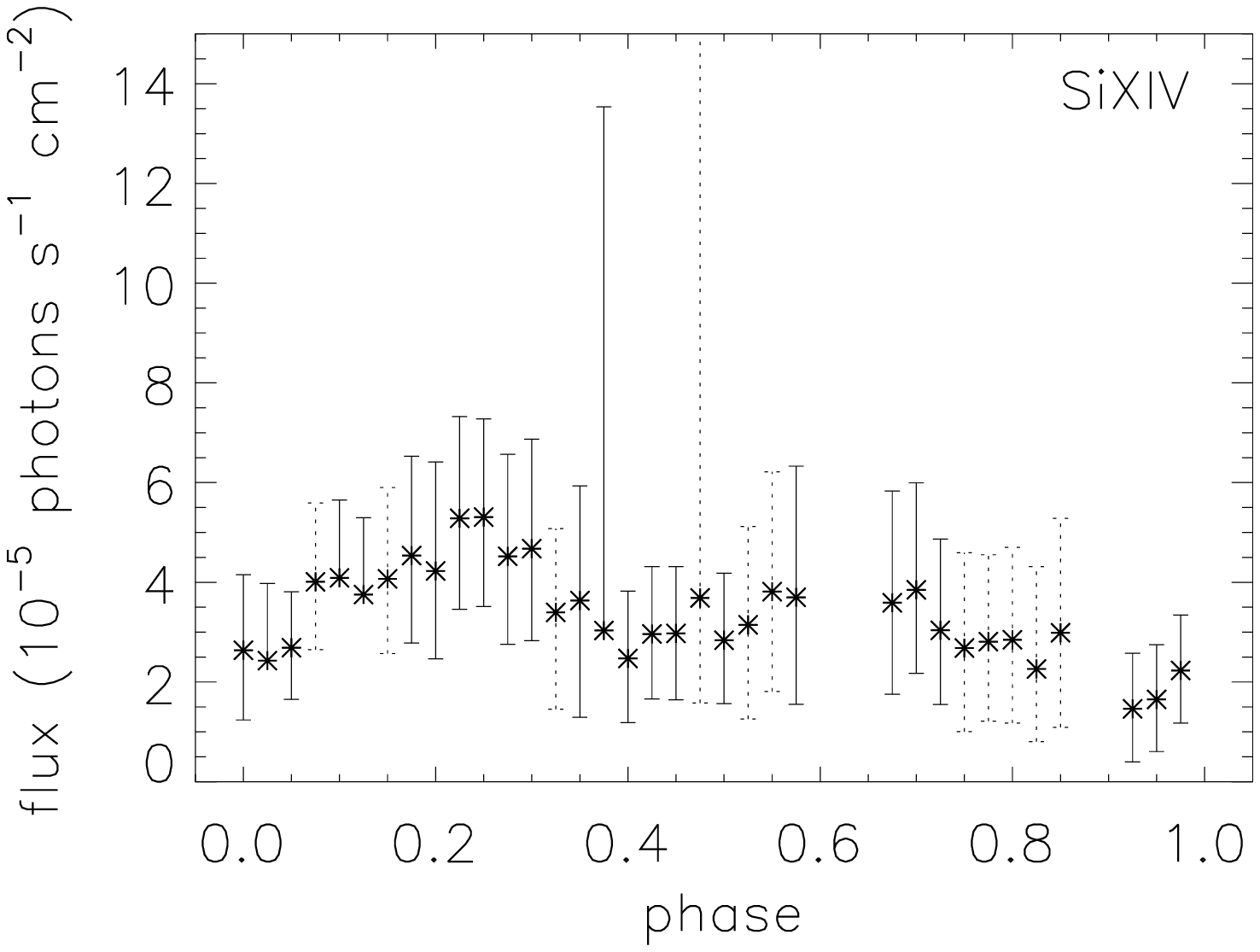}
}
\mbox{
\includegraphics[width=2.4in]{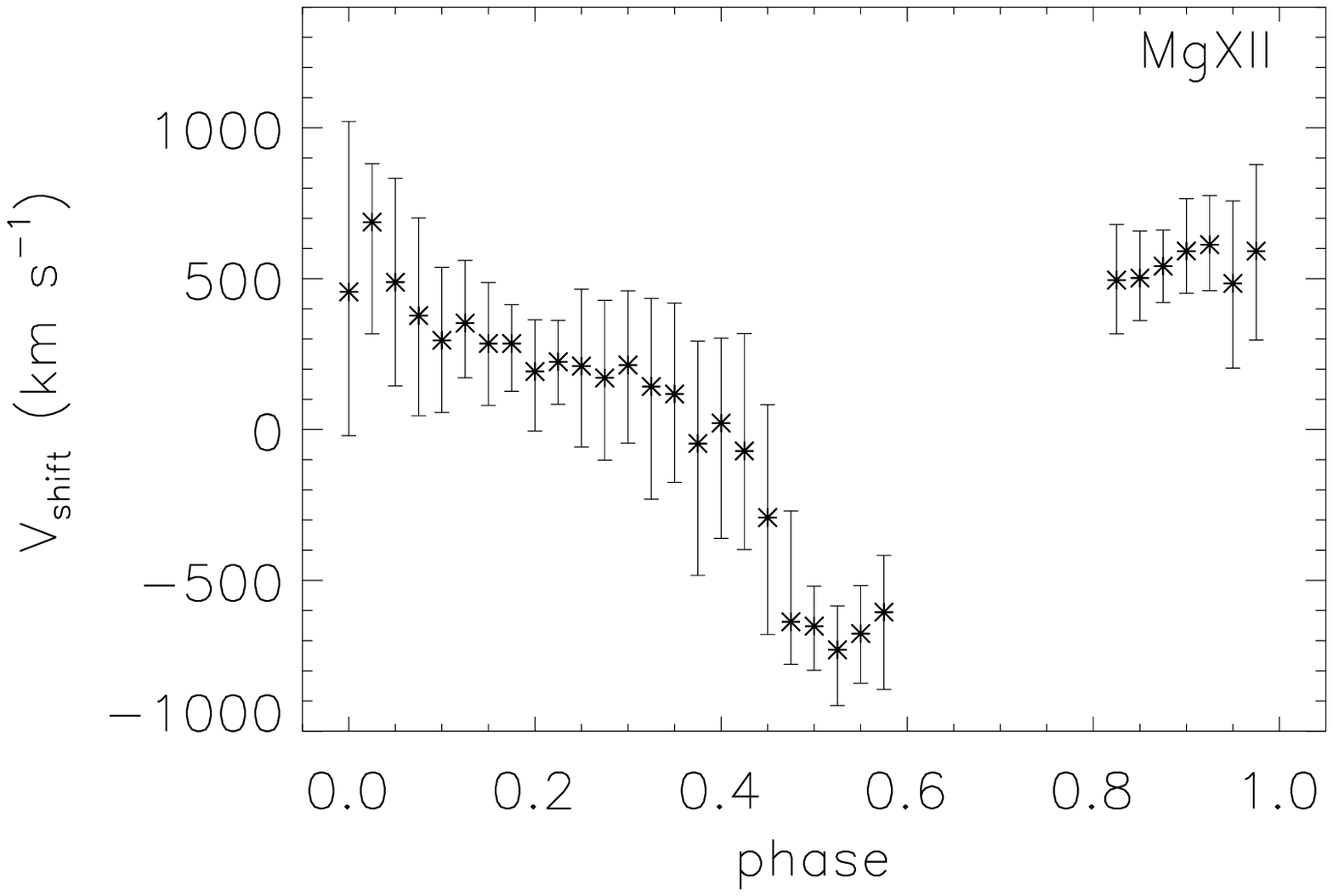}
\hskip -0.28in
\includegraphics[width=2.4in]{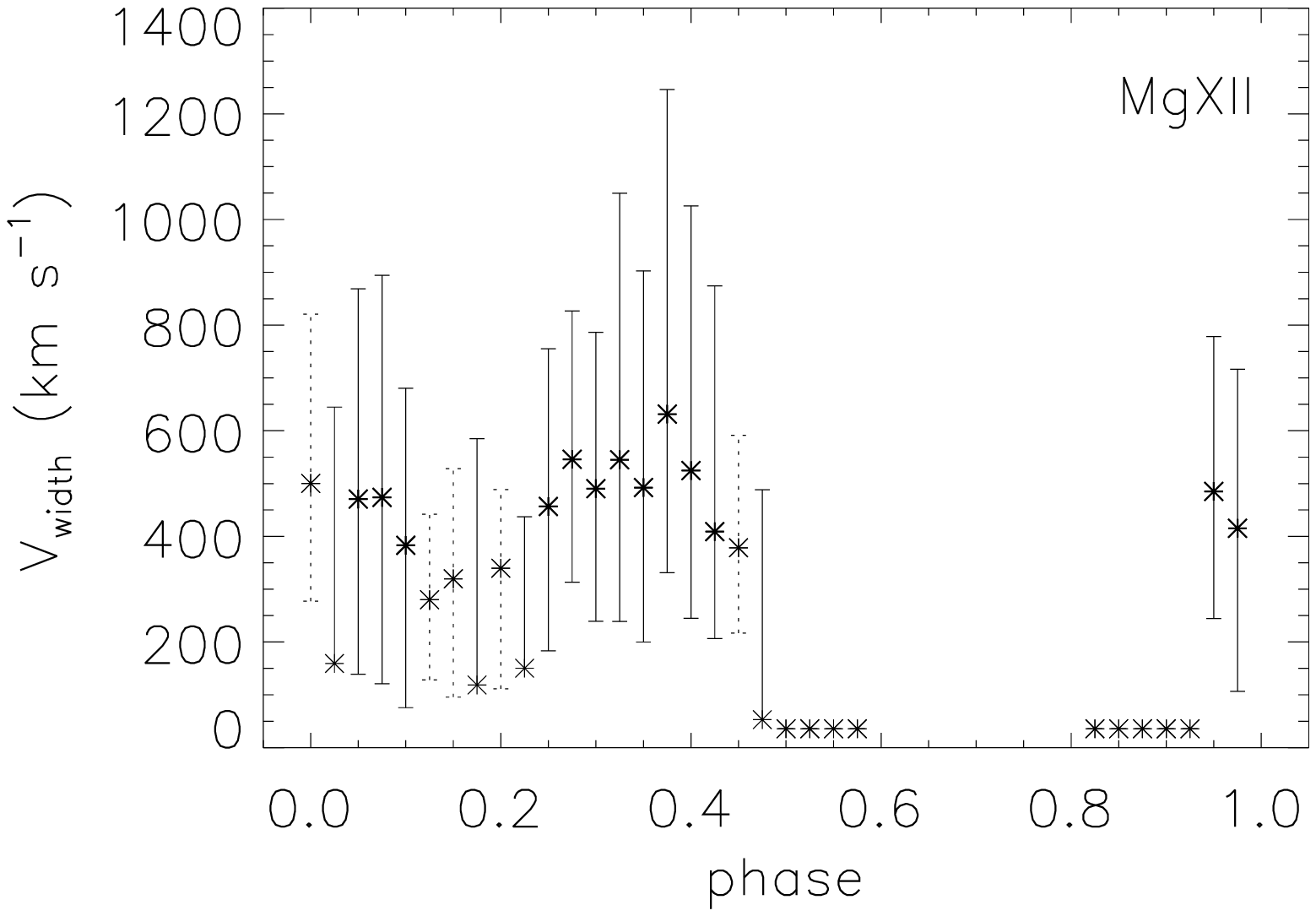}
\hskip -0.3in
\includegraphics[width=2.4in]{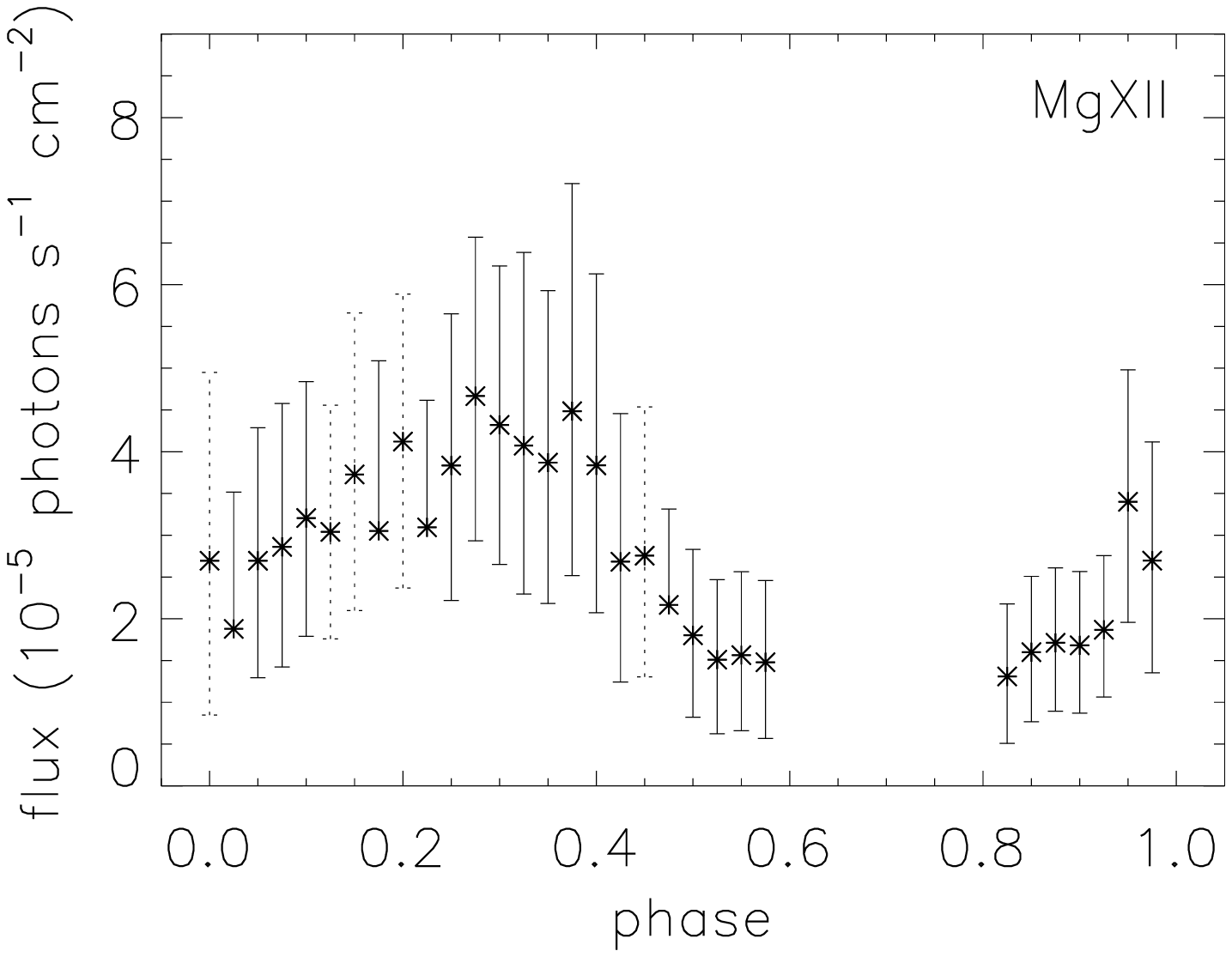}
}
\mbox{
\includegraphics[width=2.4in]{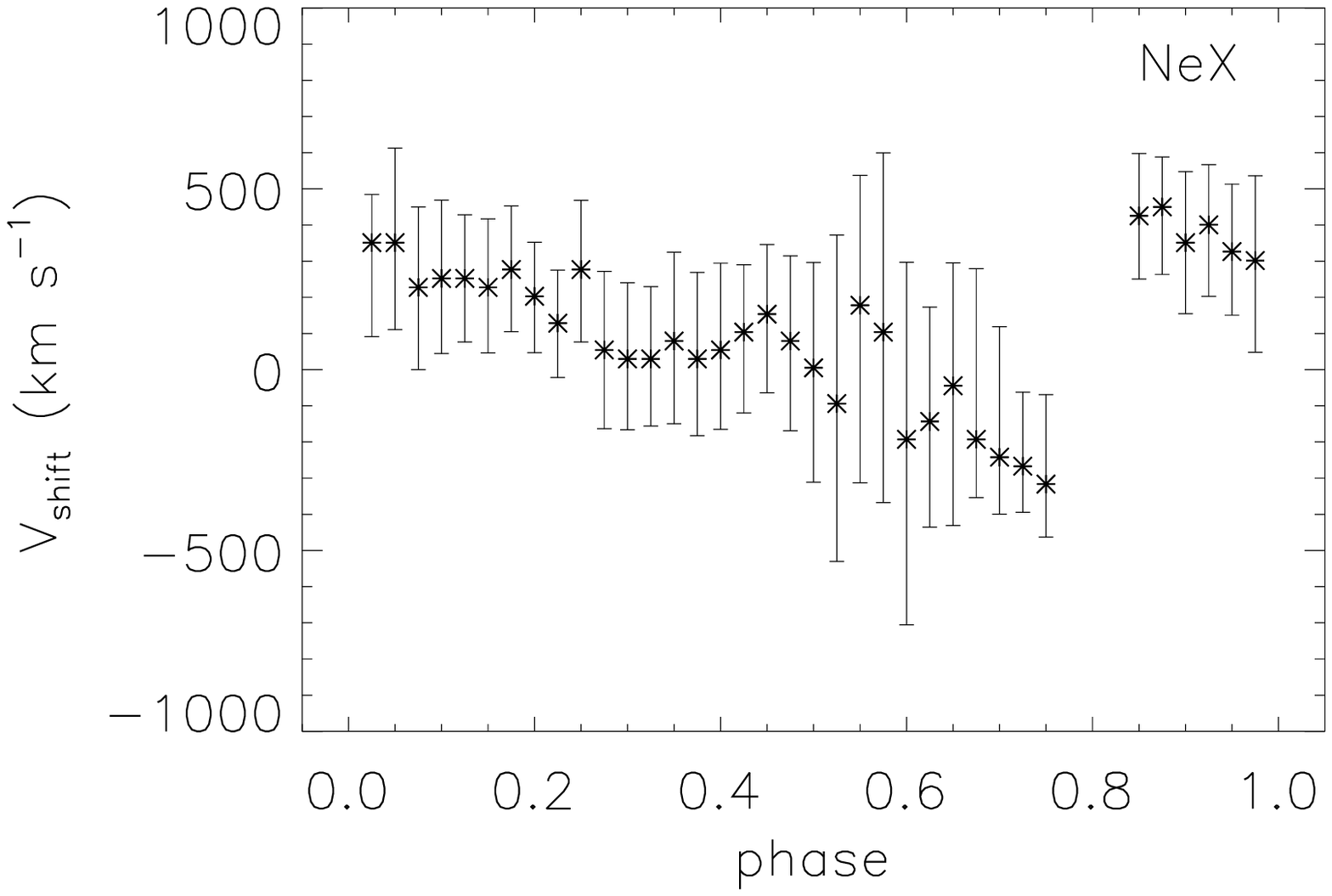}
\hskip -0.28in
\includegraphics[width=2.4in]{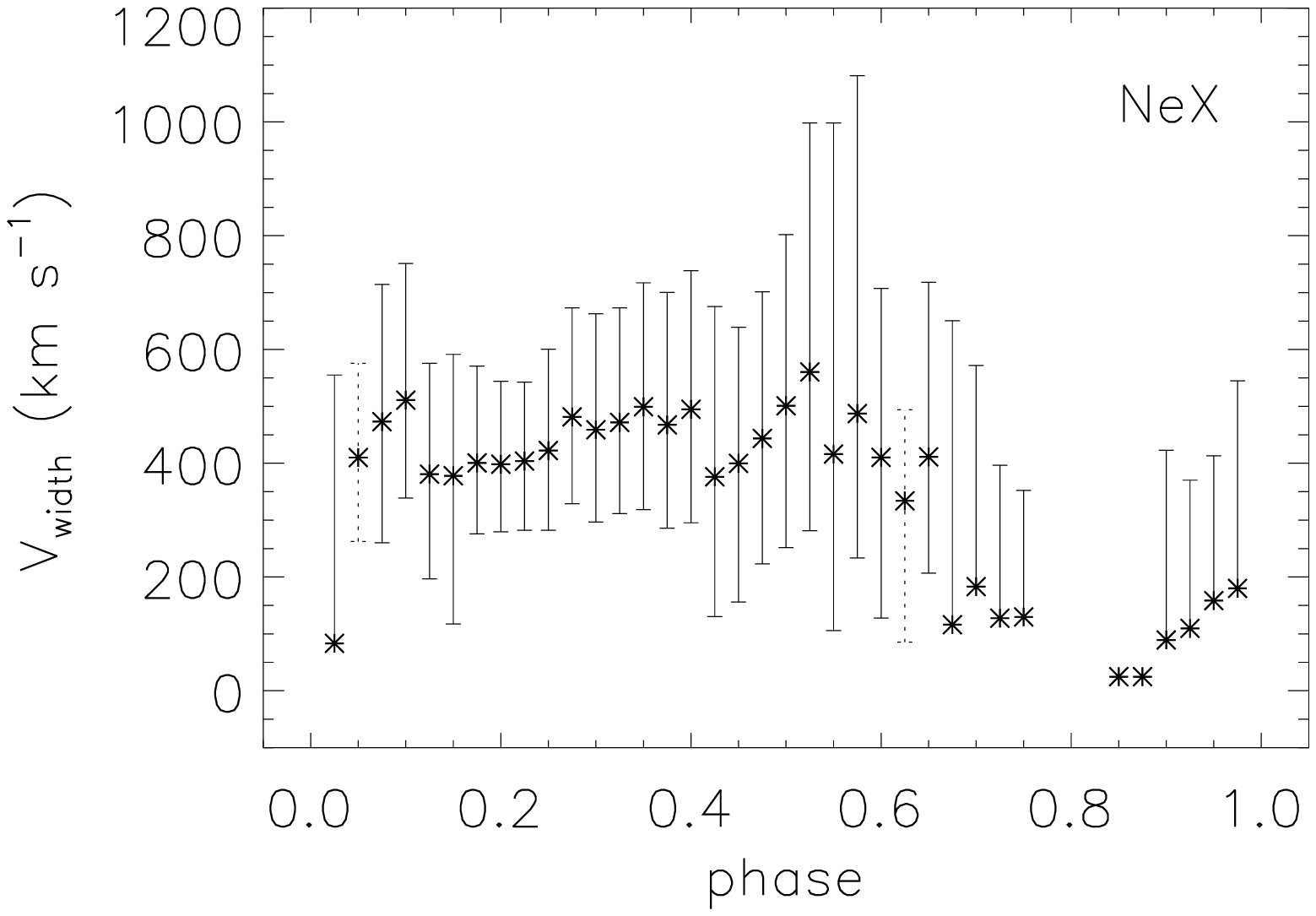}
\hskip -0.3in
\includegraphics[width=2.4in]{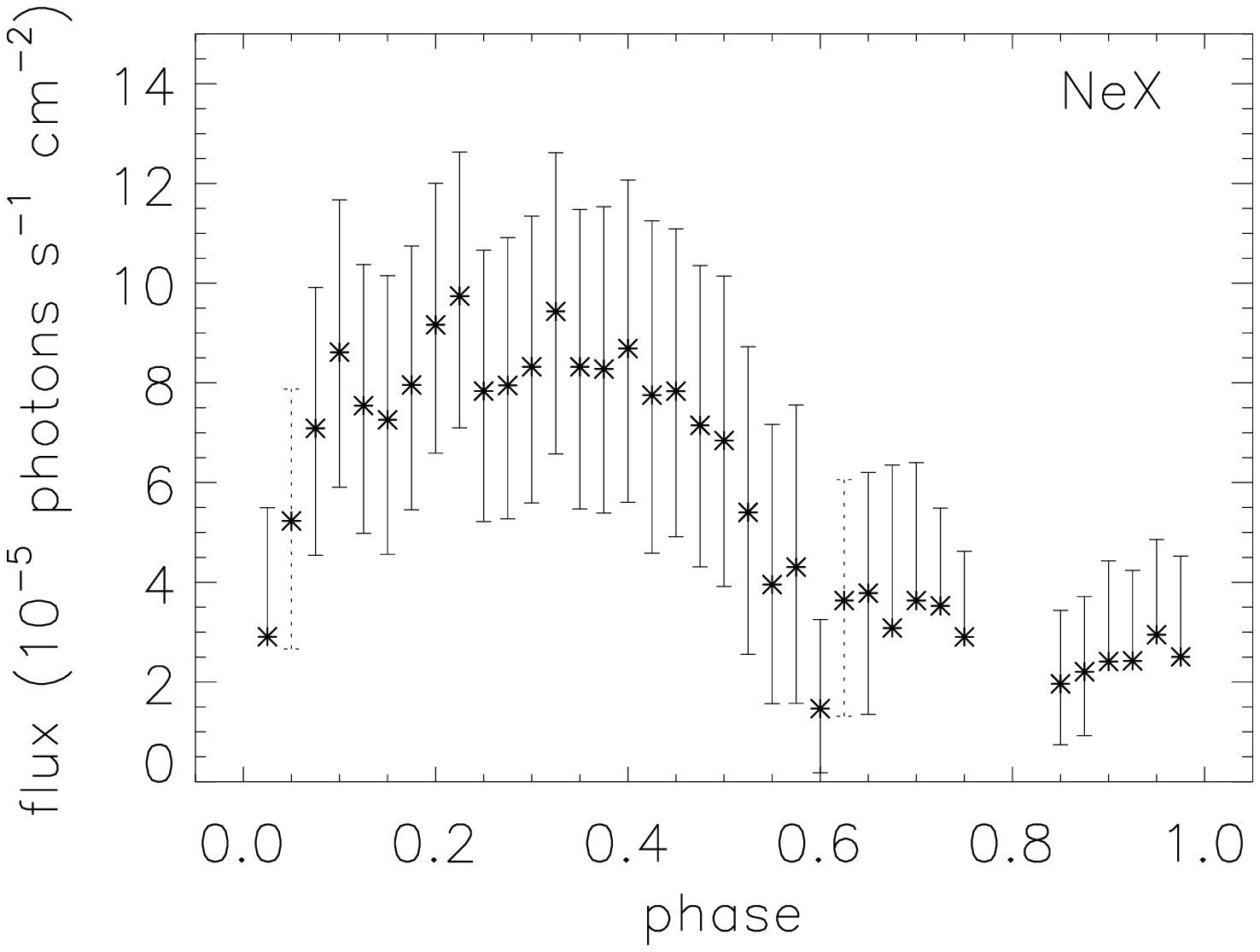}
}

\mbox{
\includegraphics[width=2.4in]{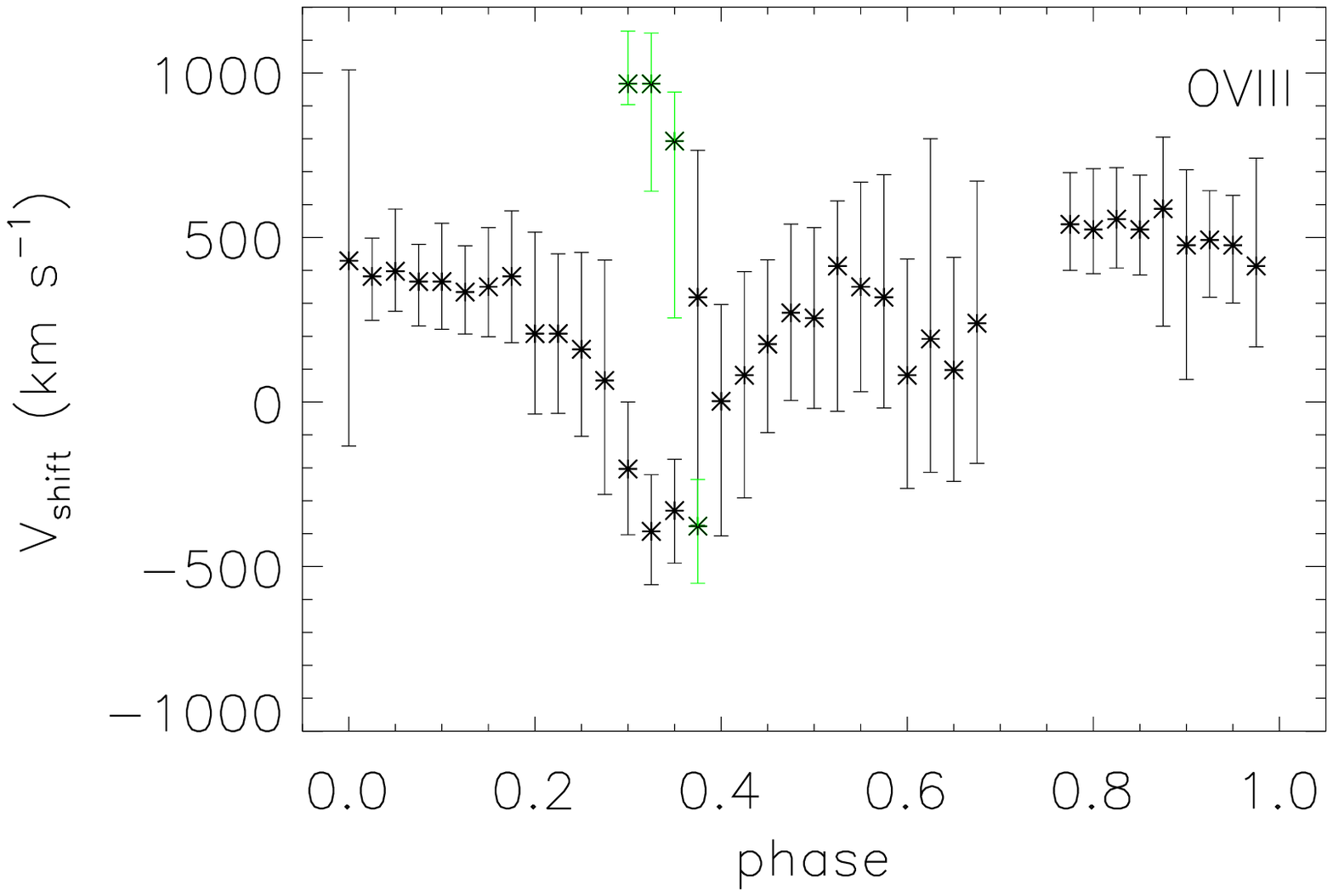}
\hskip -0.28in
\includegraphics[width=2.4in]{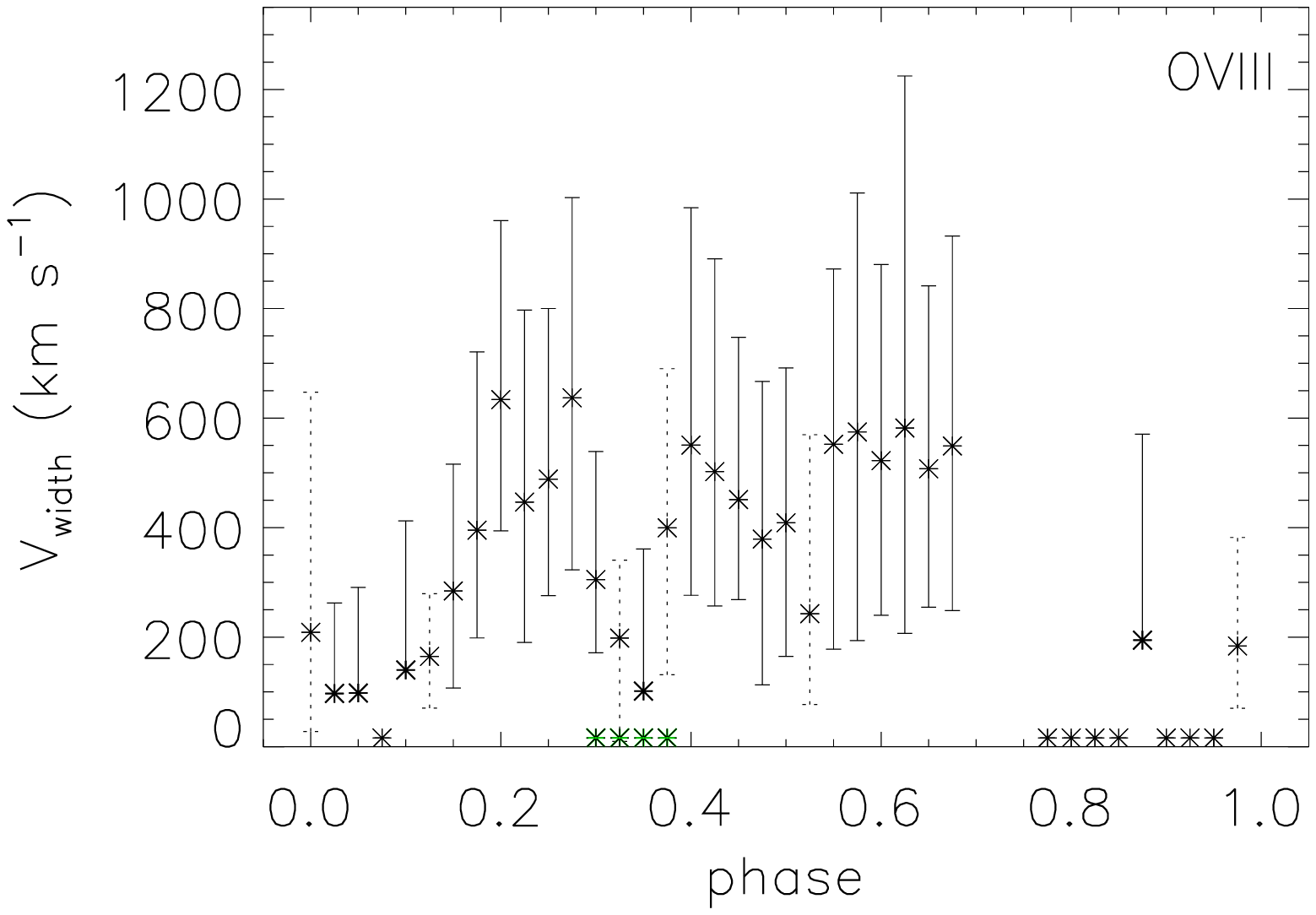}
\hskip -0.3in
\includegraphics[width=2.4in]{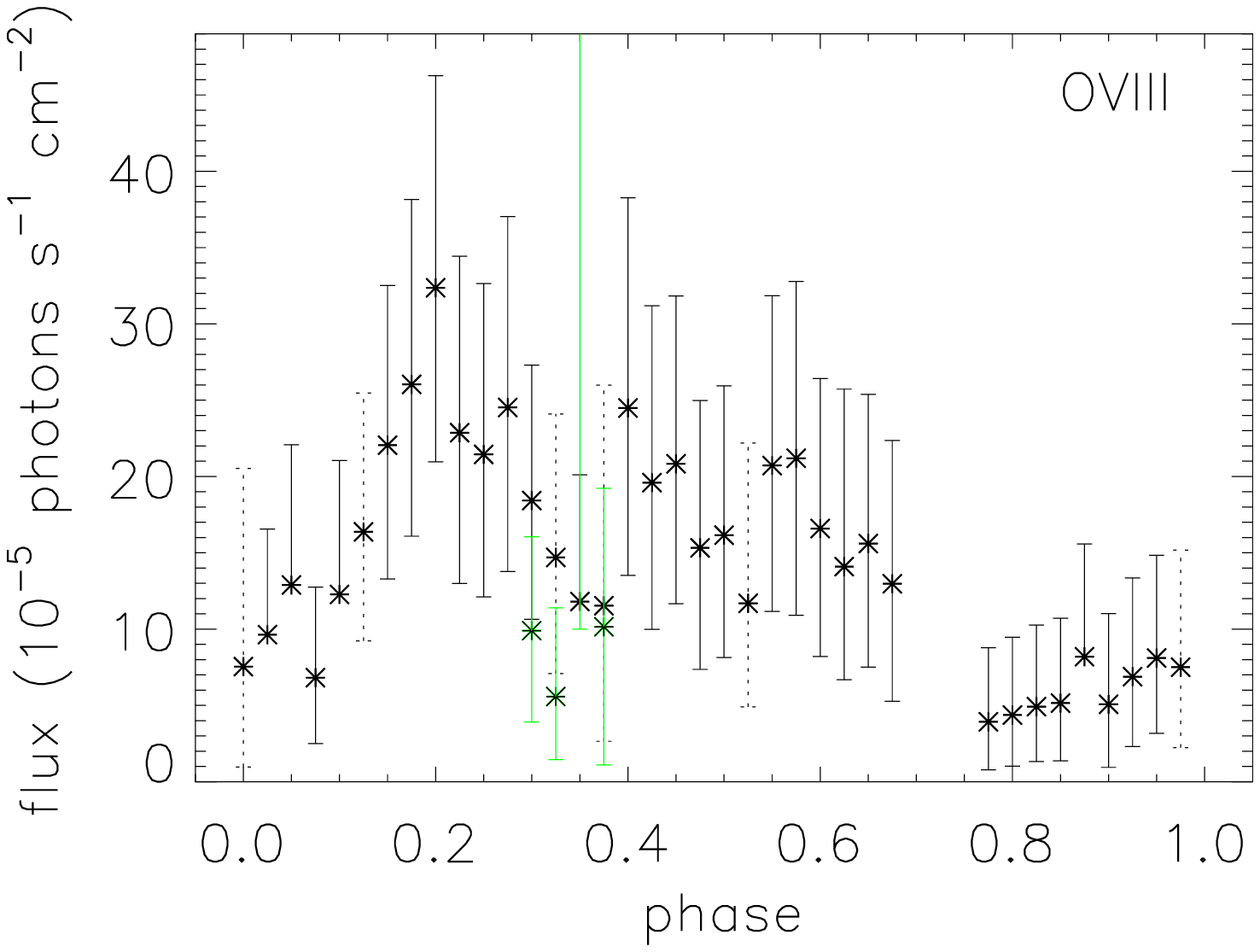}
}

\caption{ Line red-shift (left), broadening (middle), and flux (right)
  as a function of phase for Ly $\alpha$ lines of \sixiv, \mgxii,
  \nex, and \oviii~ with $3\sigma$ (solid line) and $1\sigma$ (dotted
  line) error bars shown. Line broadening is fixed at 0.001\AA\ for
  those points without error bars. Green lines mark additional
  detected line components.}
\label{fig-Lya}
\end{figure}

\begin{figure}
\mbox{
\includegraphics[width=2.4in]{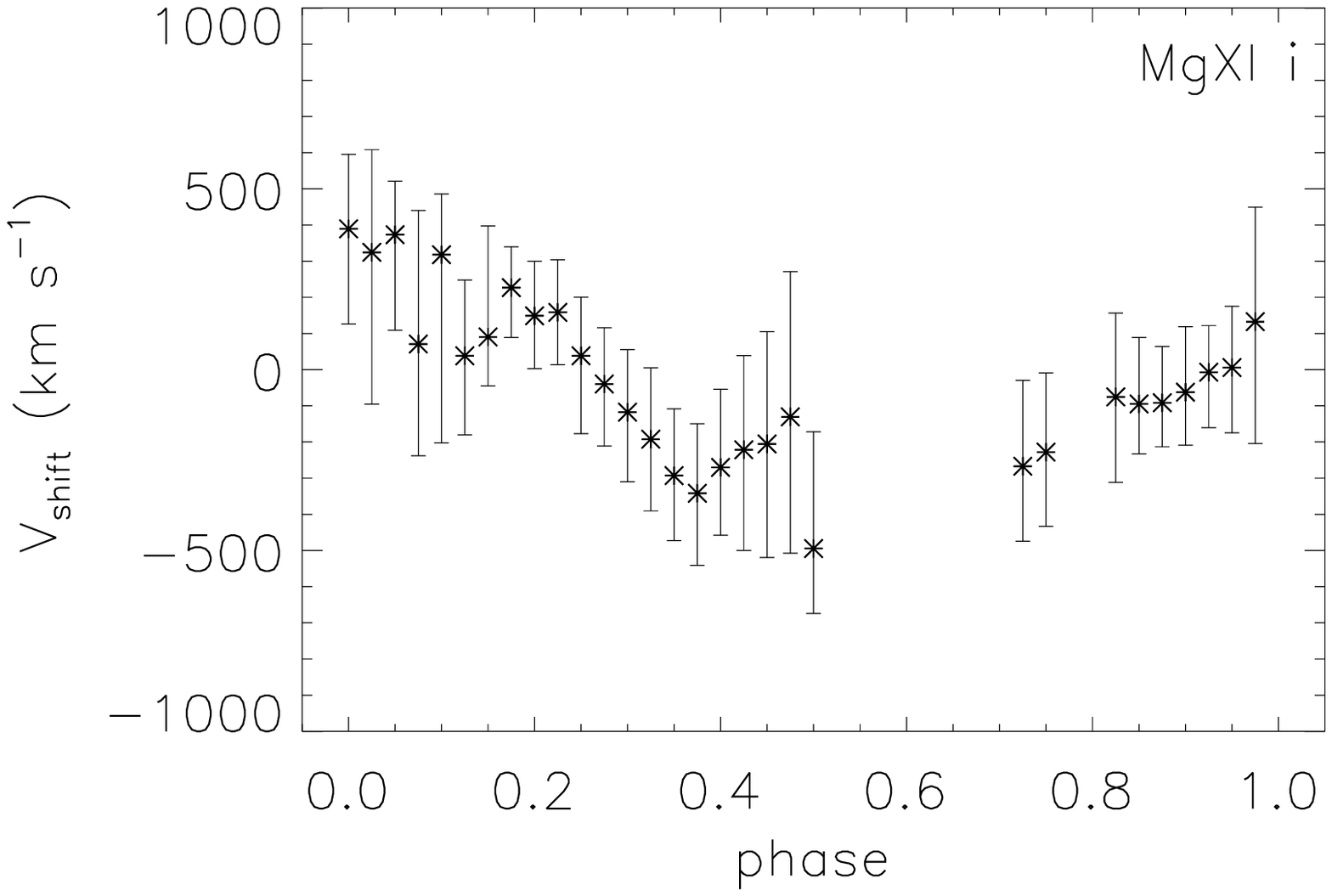}
\hskip -0.28in
\includegraphics[width=2.4in]{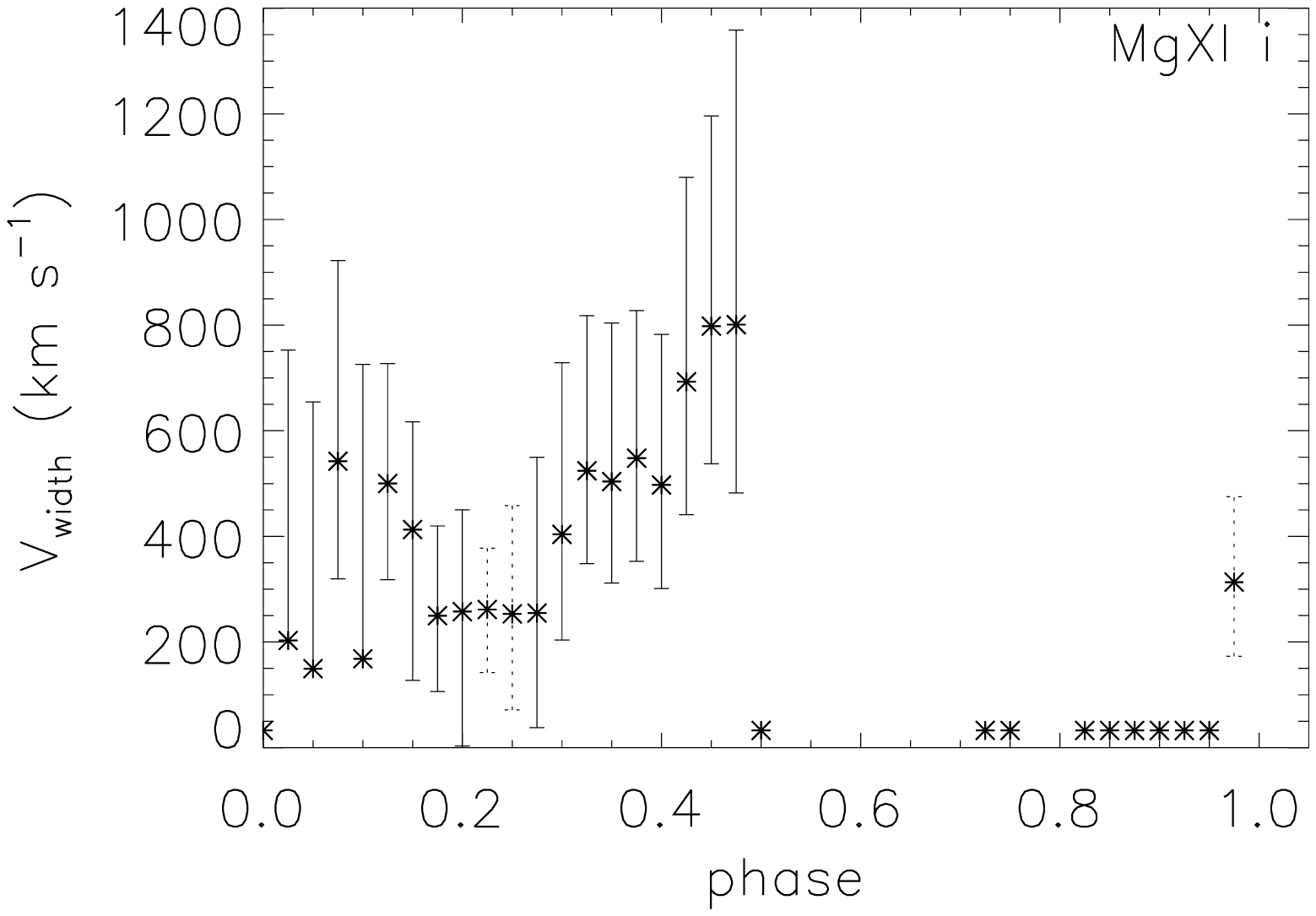}
\hskip -0.3in
\includegraphics[width=2.4in]{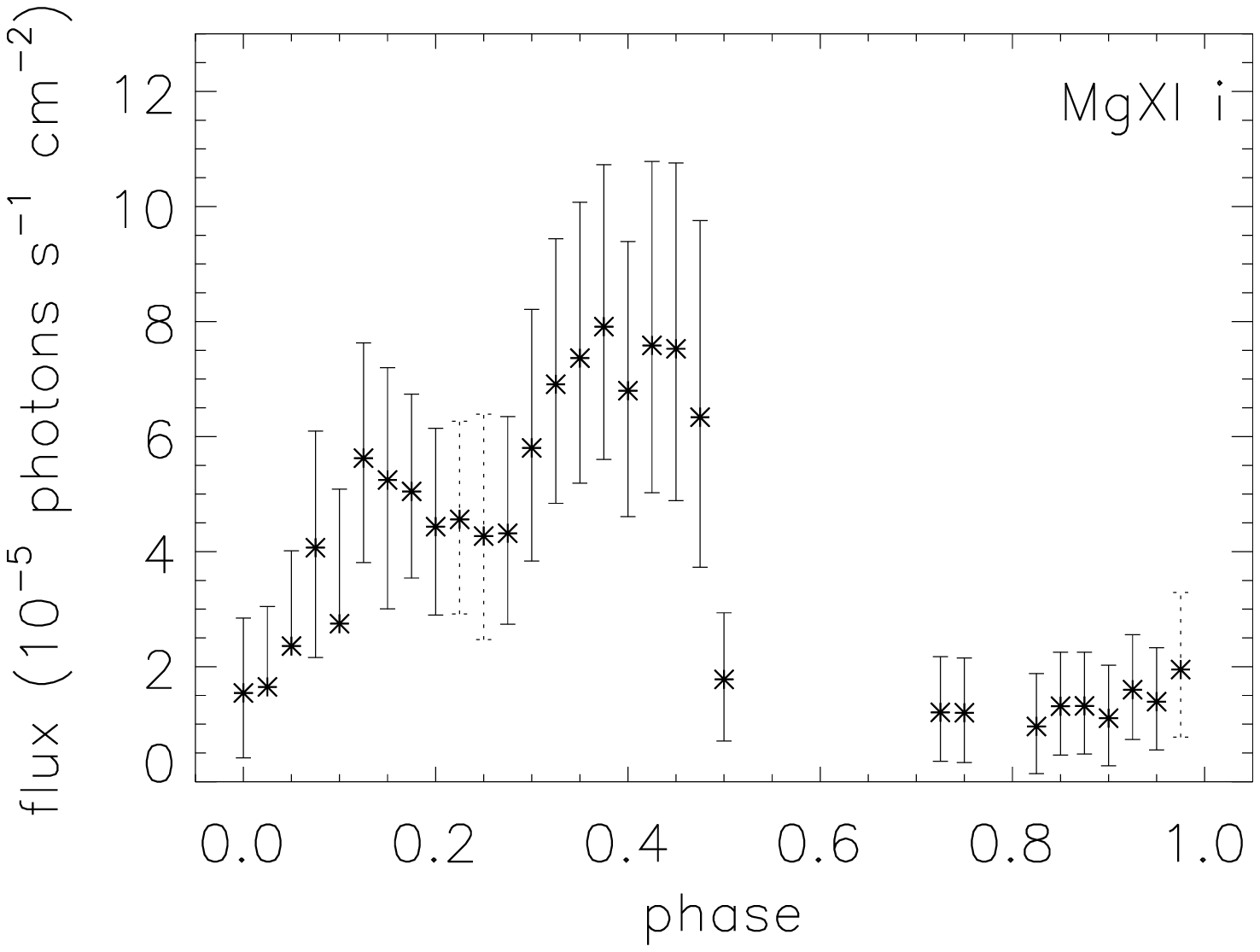}
}
\mbox{
\includegraphics[width=2.4in]{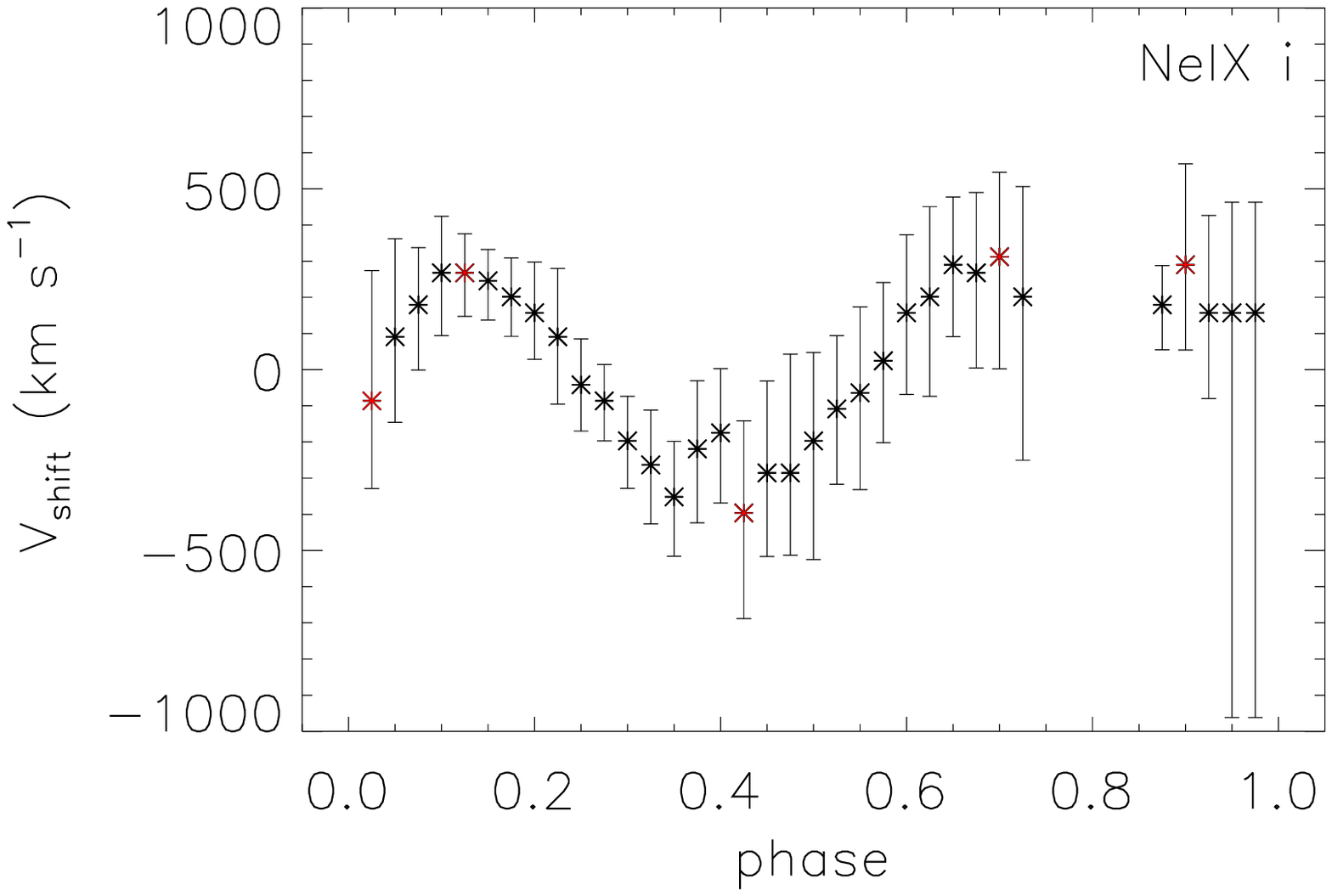}
\hskip -0.28in
\includegraphics[width=2.4in]{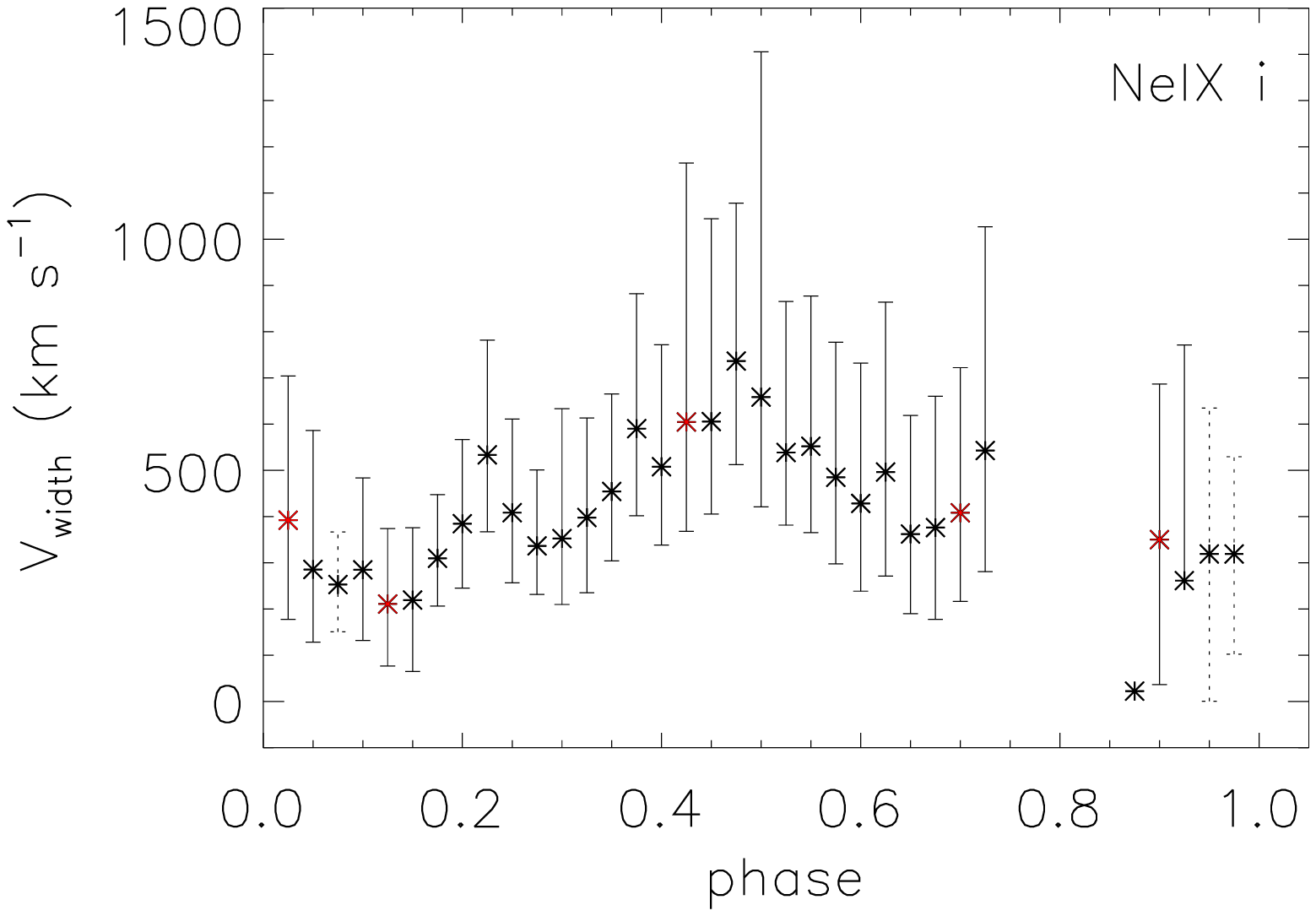}
\hskip -0.3in
\includegraphics[width=2.4in]{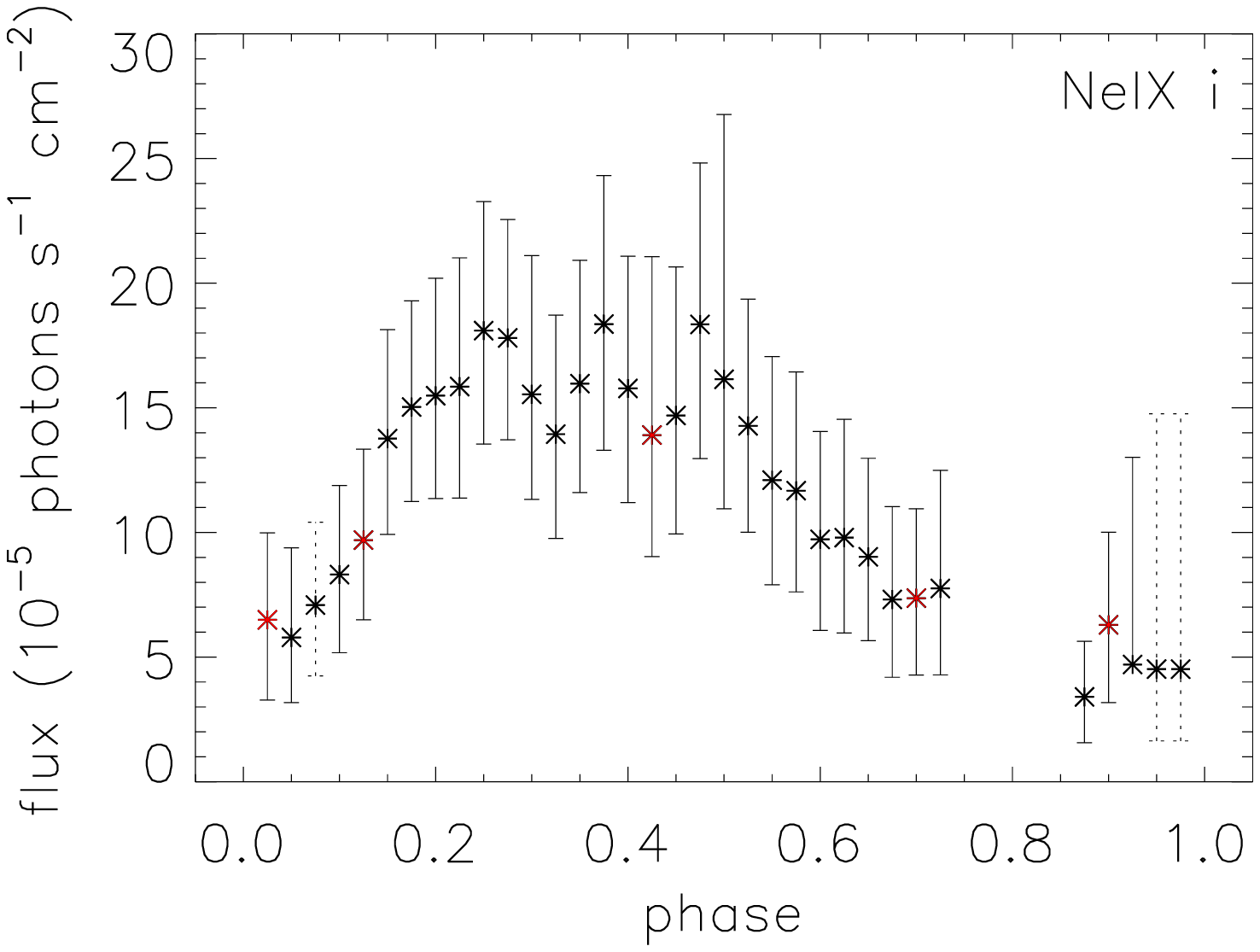}
}
\mbox{
\includegraphics[width=2.4in]{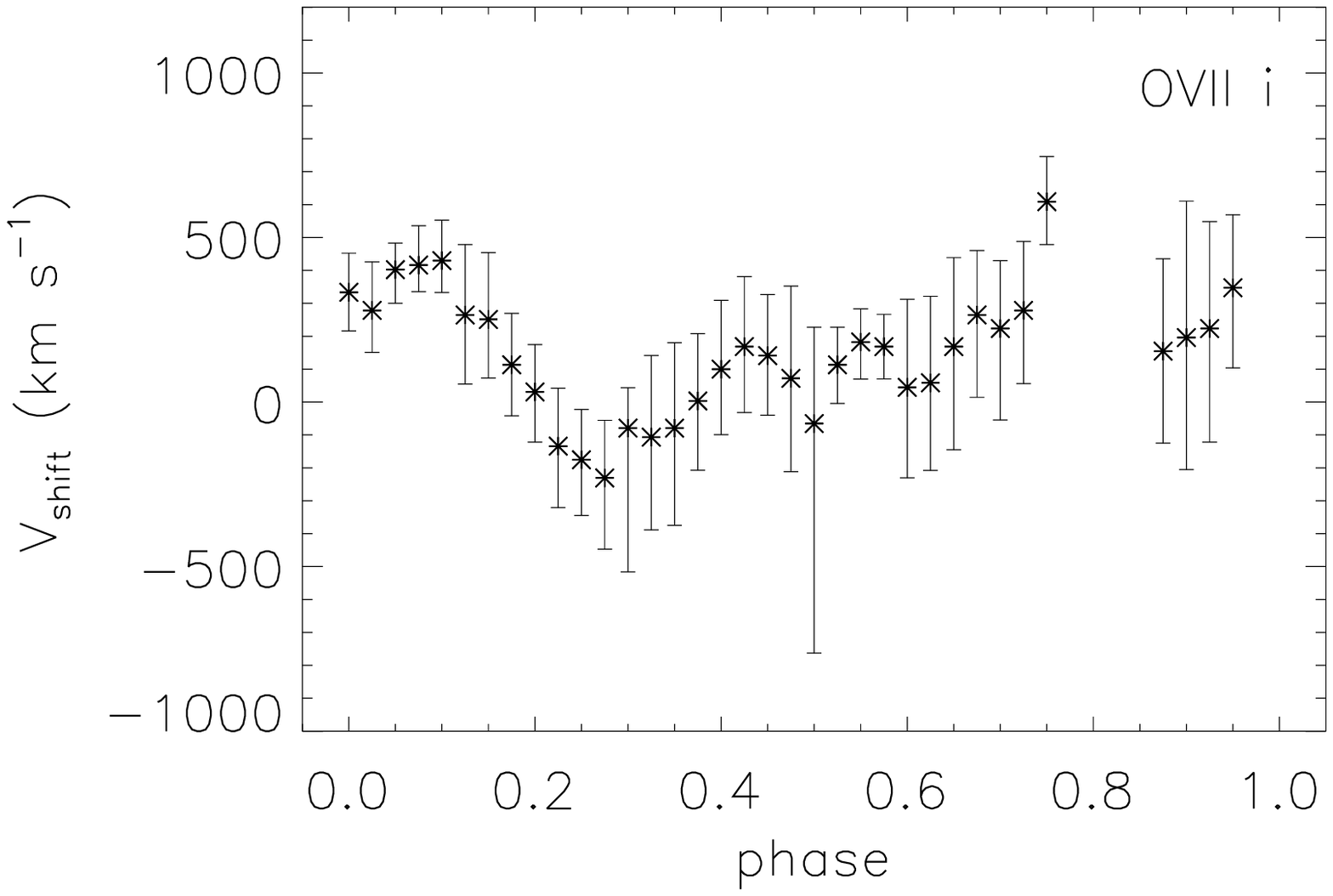}
\hskip -0.28in
\includegraphics[width=2.4in]{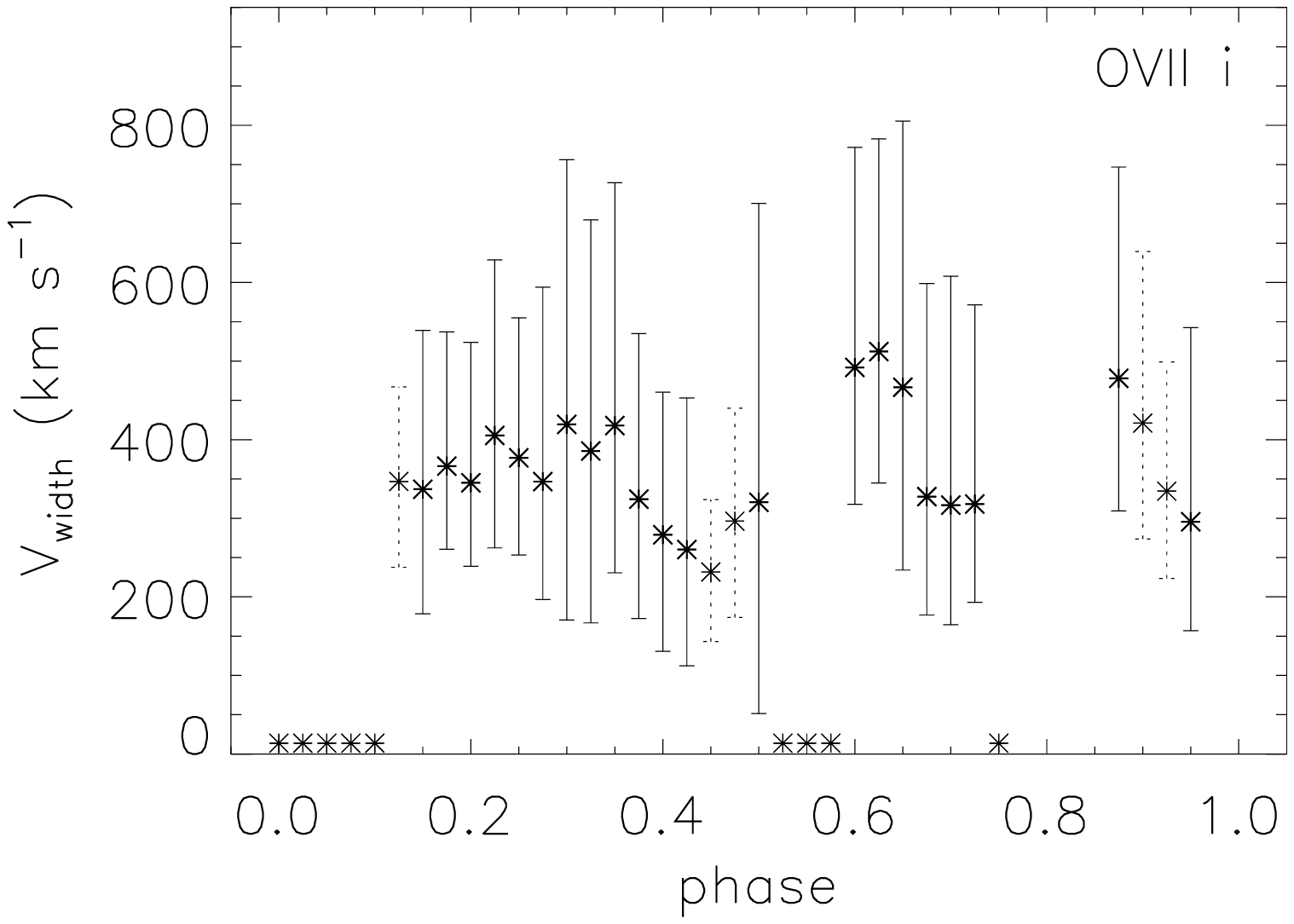}
\hskip -0.3in
\includegraphics[width=2.4in]{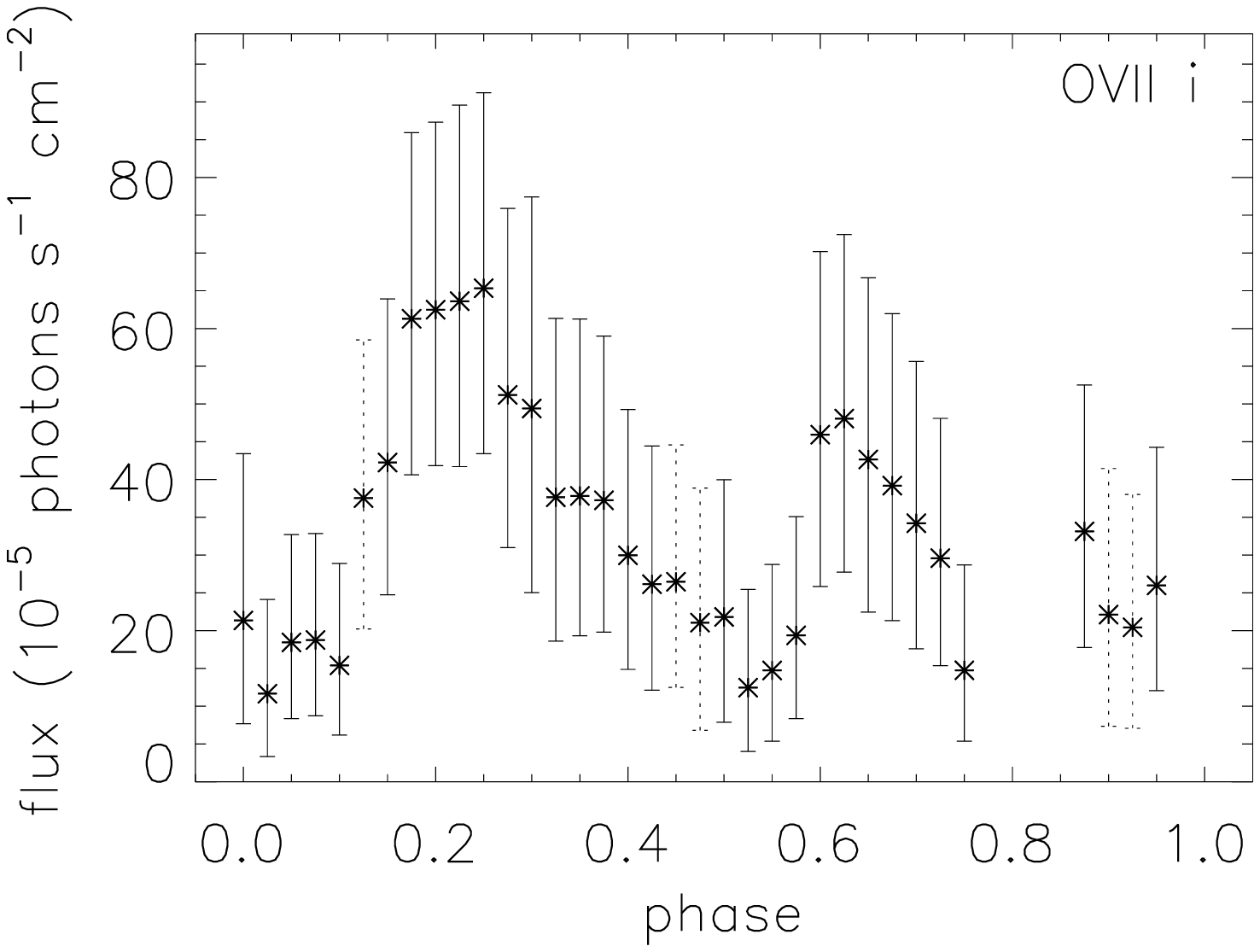}
}
\caption{ Line redshift (left), broadening (middle), and flux (right)
  as a function of phase for the intercombination lines of \mgxi~
  (upper), \neix~ (middle), and \ovii~ (lower), with $3\sigma$ (solid
  line) and $1\sigma$ (dotted line) error bars shown,
  respectively. Line broadening is fixed at 0.001\,\AA\ for those
  points without error bars. Red points mark those phases where the
  spectra are shown in Figure \ref{fig-NeIX-spec}. }
\label{fig-tri}
\end{figure}

The preliminary analysis in \citet{cottam2001} only had two binary
orbits at hand, which provided four phase bins and tentative
suggestions with respect to an orbital phase dependence of bright line
centroids. Incorporating the seven orbits in the new observations
allows for a much more detailed analysis of the line phase dependence.
These line properties include line centroids, line widths, and line
fluxes.  The results are shown in Figures \ref{fig-Lya} (for
Ly$\alpha$ lines) and \ref{fig-tri} (for intercombination lines) and
Tables 3 and 4.  These are the brightest lines in the X-ray spectrum.
There are other weaker lines present in the spectrum; however, we
perform the phase dependent analysis only for the bright lines.

\begin{figure}
\mbox{
\includegraphics[width=2.2in]{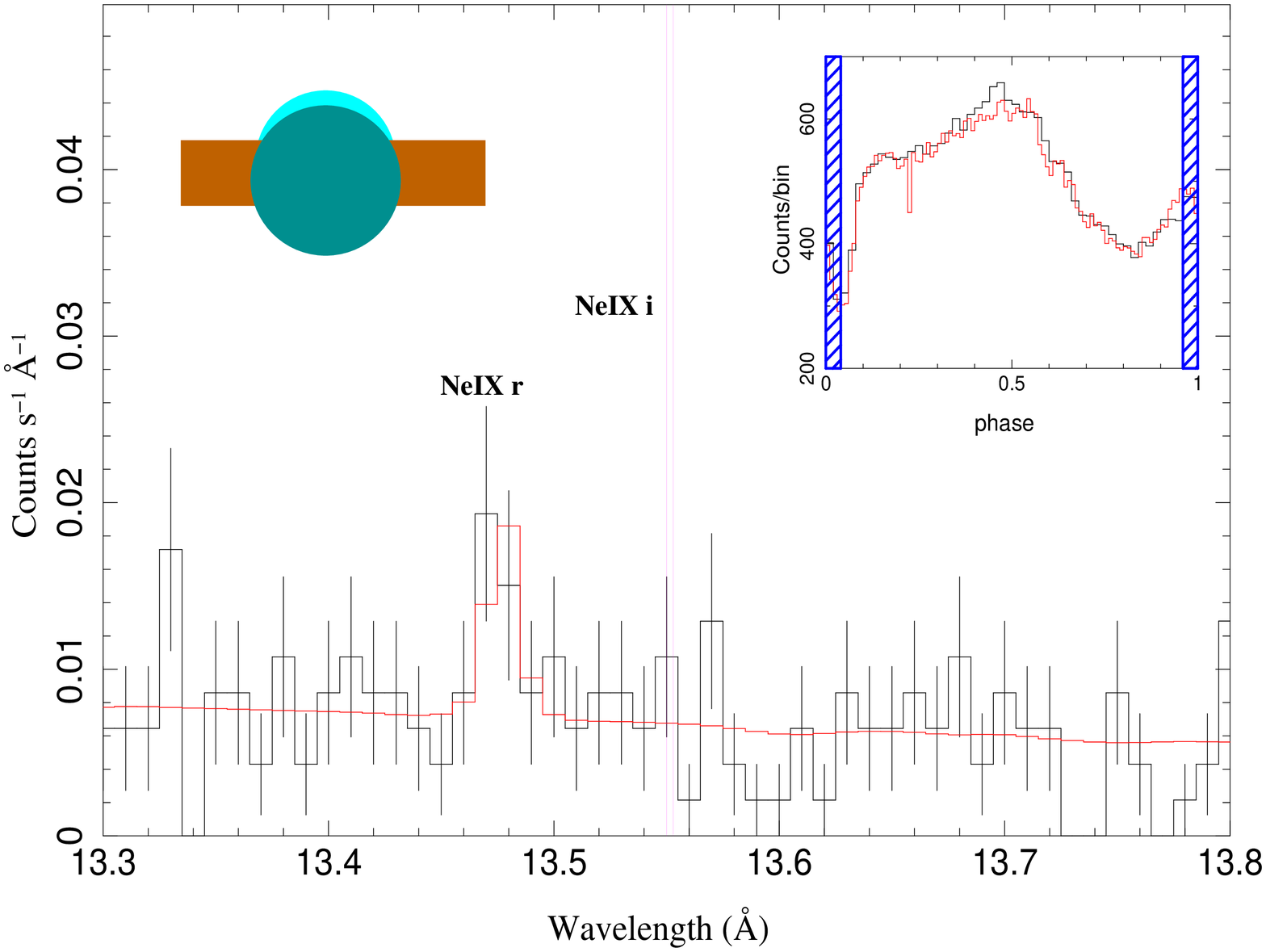}
\includegraphics[width=2.2in]{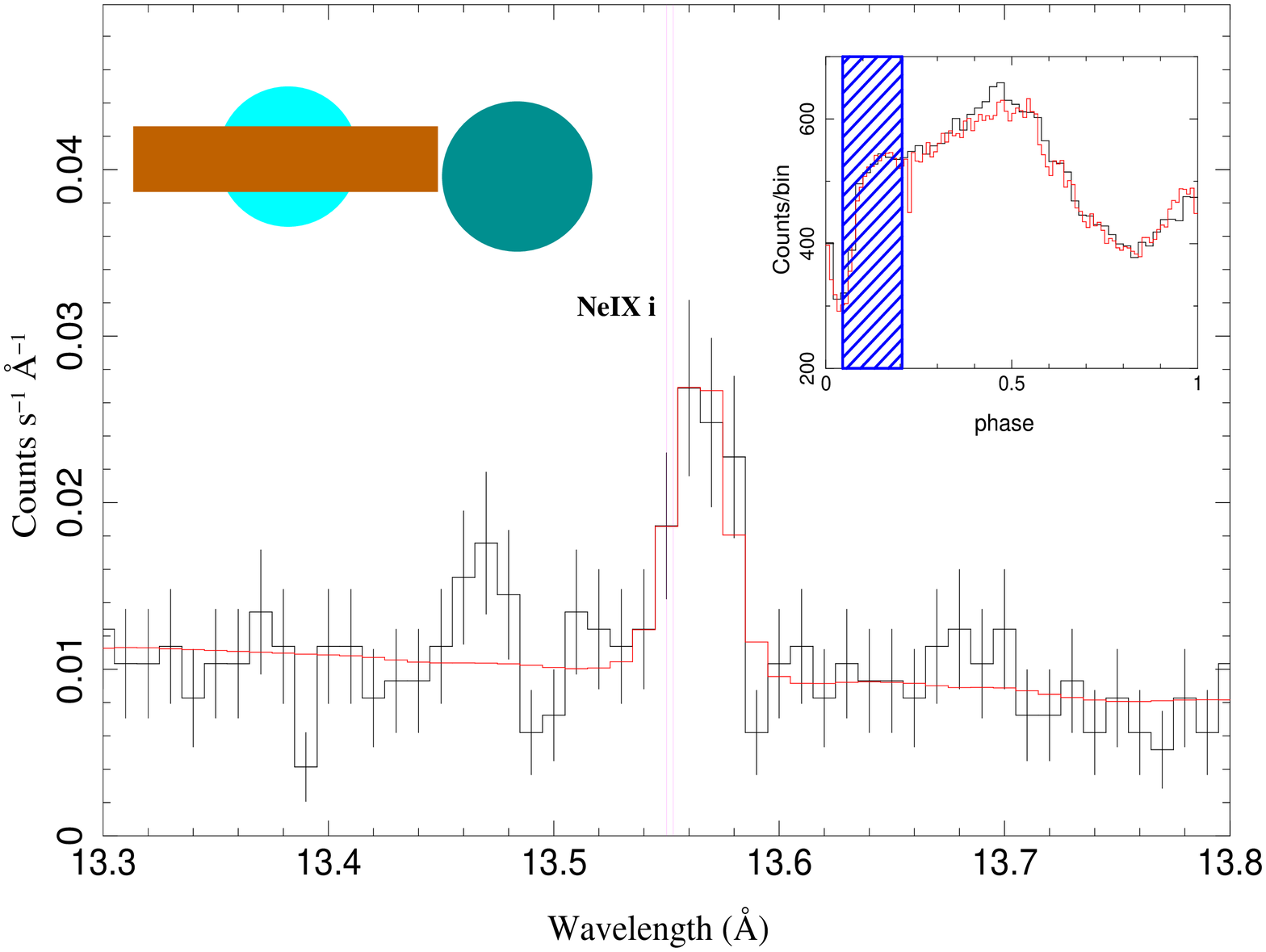}
\includegraphics[width=2.2in]{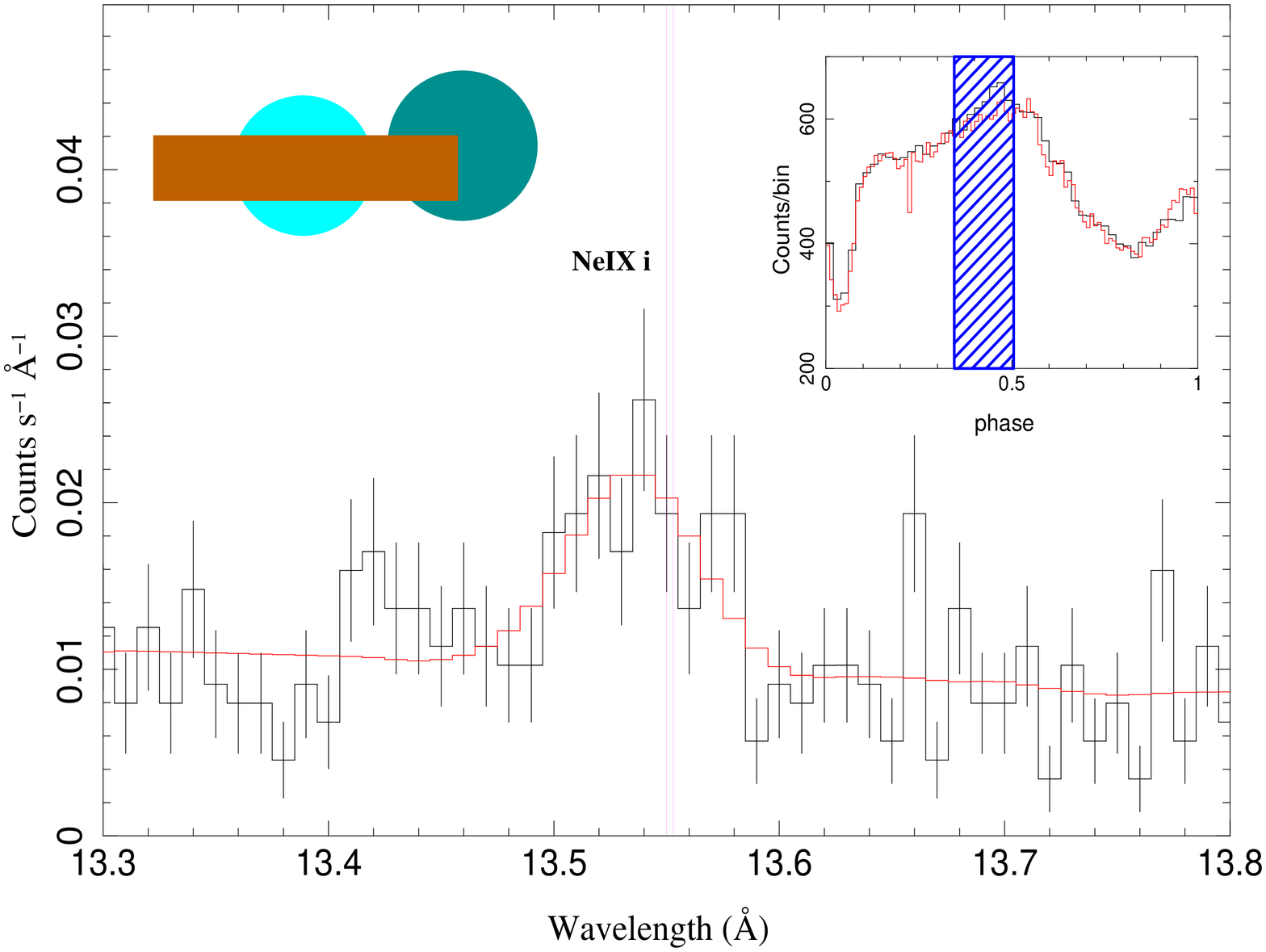}
}
\mbox{
\includegraphics[width=2.2in]{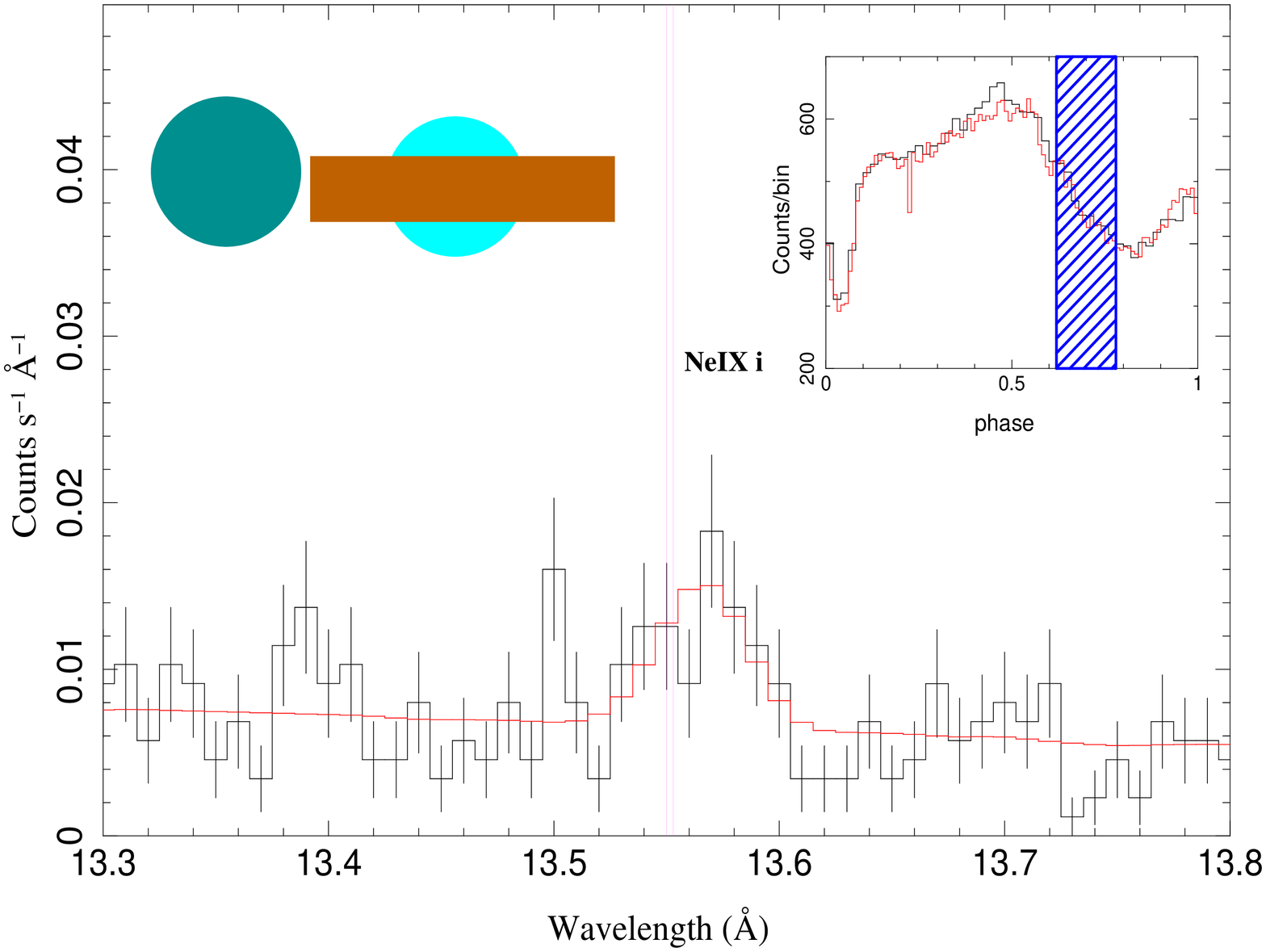}
\includegraphics[width=2.2in]{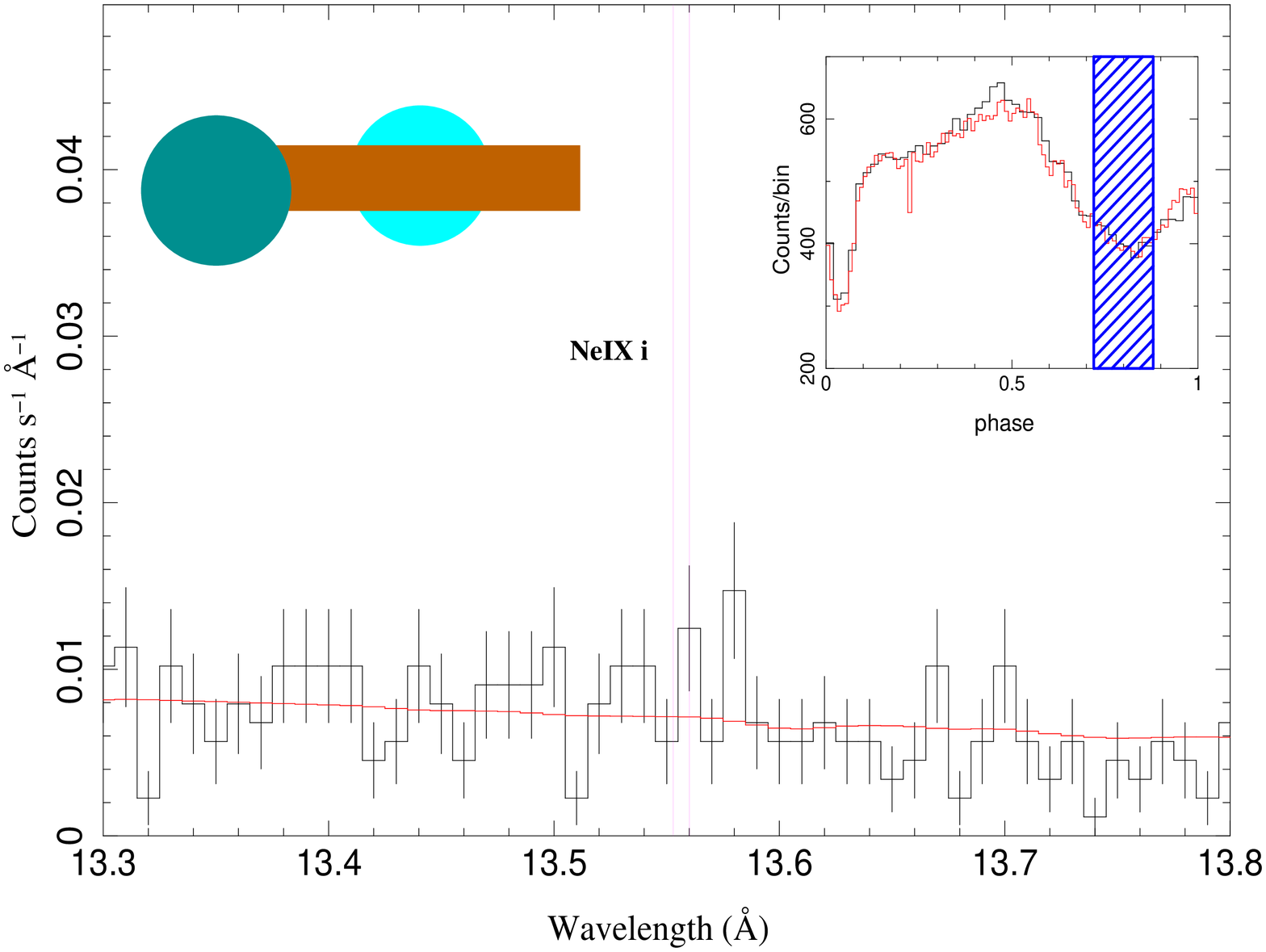}
\includegraphics[width=2.2in]{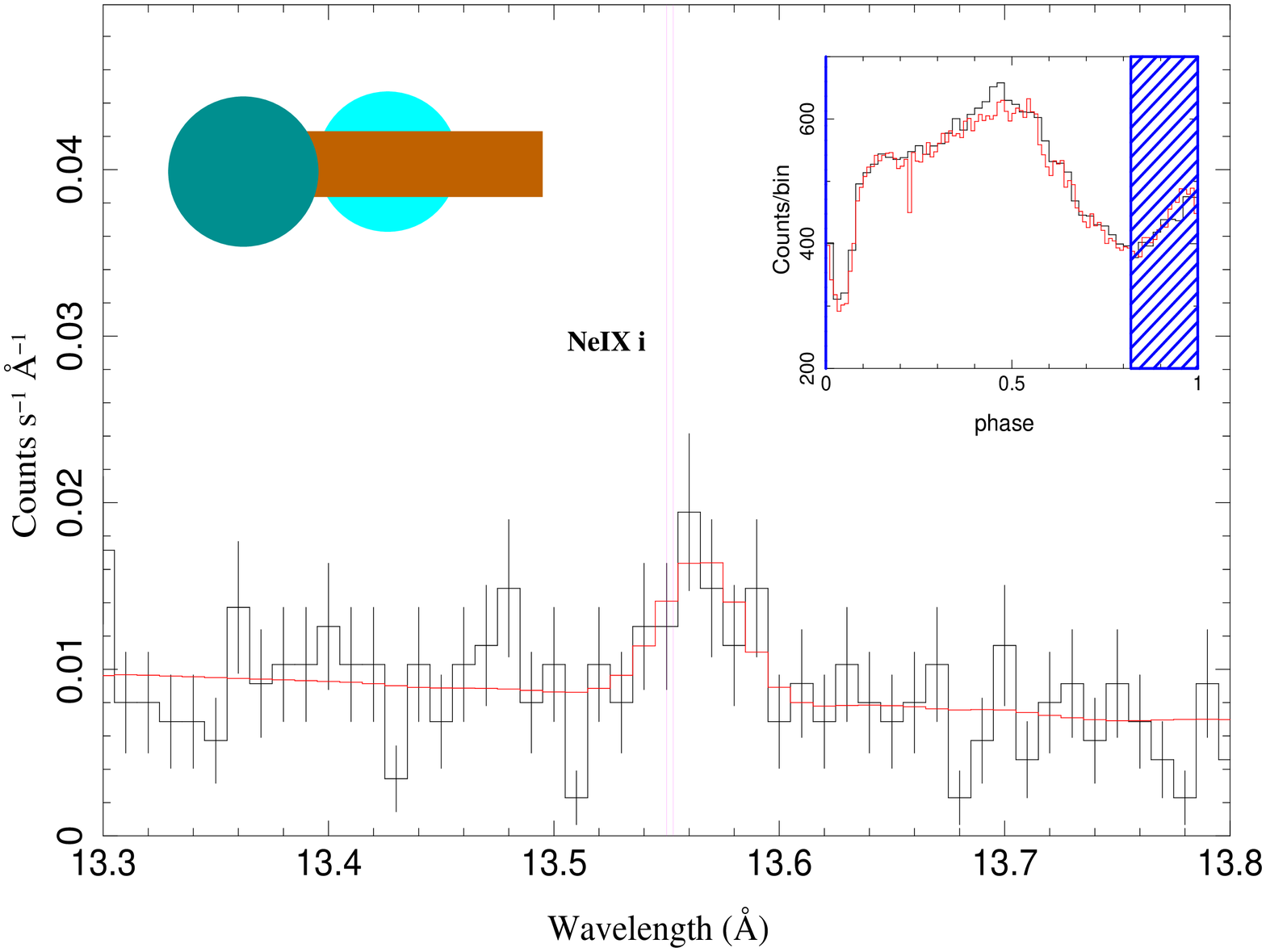}
}
\caption{ Variations of \neix~ intercombination line at the phases
  chosen in Figure \ref{fig-tri}, including the spectra without line
  detection for eclipse (left upper) and phase centered at 0.825
  (middle lower).  The pink lines mark the zero-velocity line
  location.  The cartoon shows the hypothesized structure of the
  system (based upon the geometrical model of
  \citealt{heinz2001}). Each spectrum represents the average for the
  phase bin outlined by the blue shadow on the light curve in the
  right corner of each figure. }
\label{fig-NeIX-spec}
\end{figure}

All of the lines show significant wavelength shifts with orbital
phase. Most lines follow a very similar shift pattern. At phase 0 the
lines appear highly red-shifted and exhibit values between 400--$500
{\rm km~s}^{-1}$.  These values decline to zero near about phase 0.25
and turn into increasing blue-shifts of up to 400--$500 {\rm
km~s}^{-1}$ near phase 0.5 where in most cases the shifts reverse
again. The most prominent lines showing this pattern are \neix, \mgxi,
\mgxii, and \sixiv. In \nex the shift reversal happens later at phase
0.8, at \ovii and \oviii~ somewhat earlier than mid-phase. The wave
pattern appears most clearly in \neix, \mgxii, and \sixiv, which are
also the brightest in the sample. Uncertainties vary between 100 and
250 $\rm km~s^{-1}$ depending on the phase and the detected flux. This
pattern is very consistent with a spatially confined emission region
moving with orbital phase.  We illustrate this with a cartoon
structure of the system and the phase light curve highlighting the
phase bin location as well as the line spectrum for the case of \neix~
in Figure \ref{fig-NeIX-spec}. Shown phases include 0, 0.1, 0.42,
0.68, 0.83, and 0.92.

The second part in Figure \ref{fig-tri} also shows the velocity width
of the lines. Even though there appears to be some subtle variations
with phase, in most cases the widths appear rather stable over the
entire orbit. The velocity widths for all lines appear very similar
near 400 ${\rm km~s}^{-1}$ with uncertainties of around 150 ${\rm
km~s}^{-1}$ per phase bin, again depending on detected flux. The very
similar widths are consistent with a radius of
(1.14$\pm0.7)\times10^{11}$\,cm.

\begin{figure}
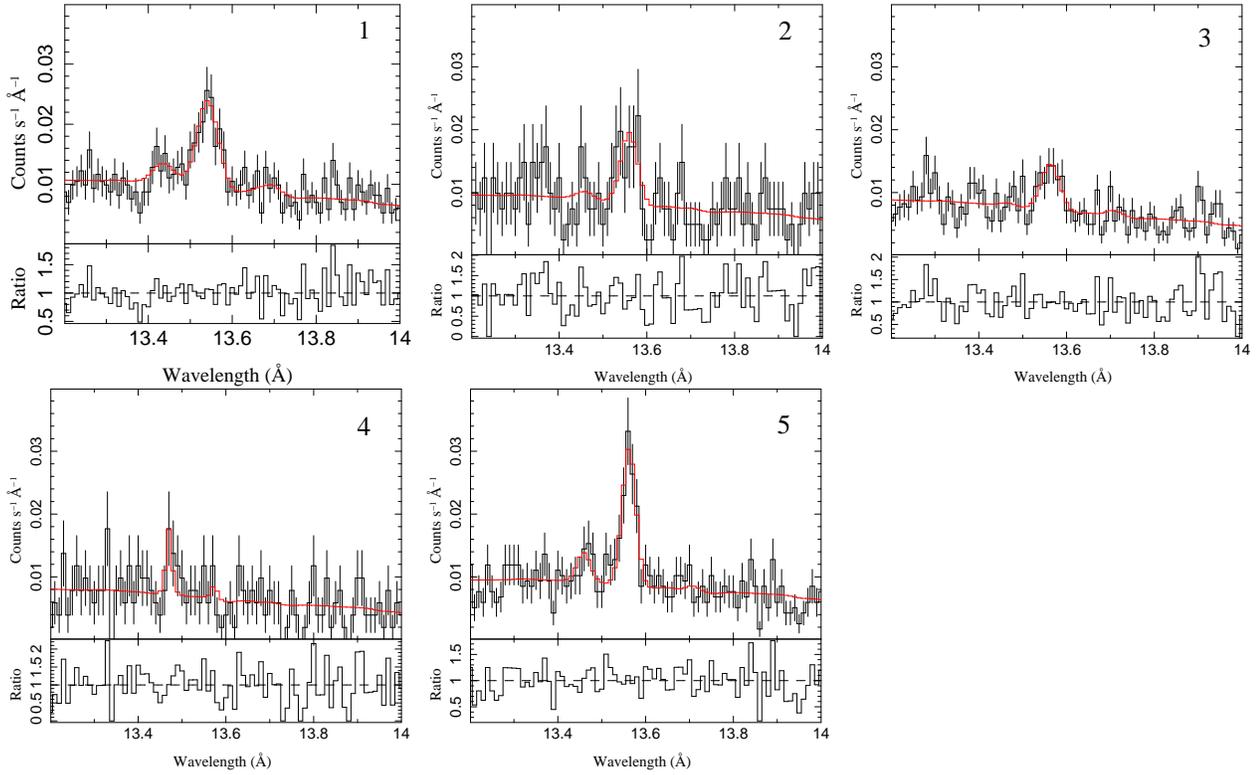

\mbox{
\includegraphics[width=2.0in,angle=270]{newobs_tri_fit_NeIX_pha_1.ps}
\includegraphics[width=2.0in,angle=270]{newobs_tri_fit_NeIX_pha_2.ps}
\includegraphics[width=2.0in,angle=270]{newobs_tri_fit_NeIX_pha_3.ps}
}
\mbox{
\includegraphics[width=2.0in,angle=270]{newobs_tri_fit_NeIX_pha_4.ps}
\includegraphics[width=2.0in,angle=270]{newobs_tri_fit_NeIX_pha_5.ps}
}
\caption{
\neix ~ triplet line spectra in the five phase bins shown in
  Figure~\protect{\ref{fig-ephemeris}}.
}
\label{fig-tri_NeIX}
\end{figure}

\begin{figure}
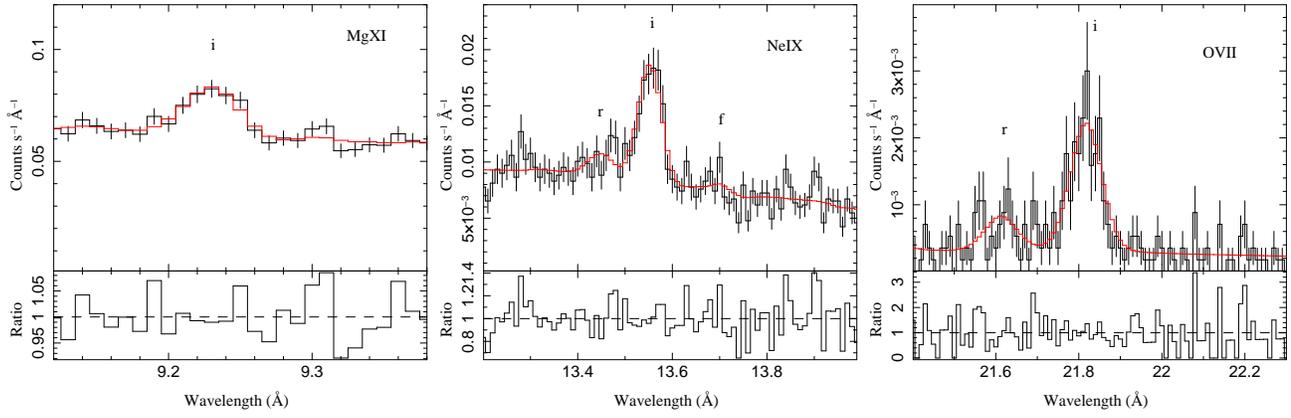

\mbox{
\includegraphics[width=0.3\textwidth,angle=270]{newobs_tri_fit_MgXI_pha_all.ps}
\includegraphics[width=0.3\textwidth,angle=270]{newobs_tri_fit_NeIX_pha_all.ps}
\includegraphics[width=0.3\textwidth,angle=270]{newobs_tri_fit_OVII_pha_all.ps}
}
\caption{Triplet line spectra \mgxi~ (left), \neix~ (middle) and
  \ovii~ (right) for the summed spectra from Obs IDs 9076 and
  9858. (See Table~\protect{\ref{fig-tri_all}}.)}
\label{fig-tri_all}
\end{figure}

Table \ref{tab-tri_all} and Figure \ref{fig-tri_all} list the results
of our detections for He-like lines of \ovii, \neix, and \mgxi. The
intercombination line flux is very bright and appears broad, while the
resonance line is weak, and we do not detect the forbidden line
component.  The resulting G ratios, defined as (i+f)/r, are all
consistent with a pure recombination plasma. The R ratios, defined as
f/i, are all very small indicating either high levels of
photo-excitation or the presence of a high density plasma.  To remove
the orbital smear we compute the R and G values for the five phase bin
case, which are shown in Figure \ref{fig-tri_NeIX} and Table
\ref{tab-tri_NeIX}. In orbital phases 2 and 4 we do not have enough
statistics to compute values, however for phases 1, 3, and 5 we
compute values and value limits which are consistent with each other.

\subsection{The Photo-ionized X-ray Spectrum
\label{photo}}

\begin{figure}
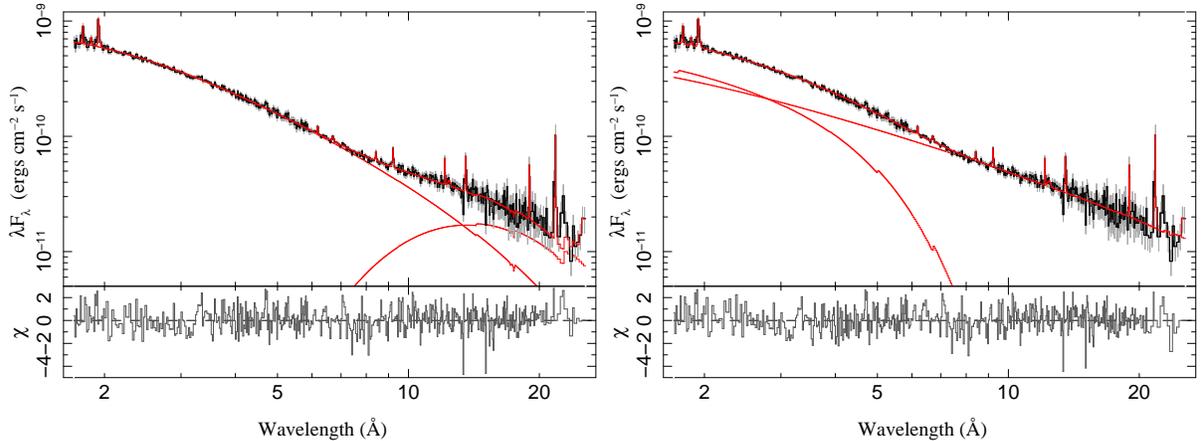

\begin{center}
\includegraphics[width=3.1in]{1822_bbody.ps}
\includegraphics[width=3.1in]{1822_partial_cover.ps}
\end{center}
\caption{Continuum fits to the X-ray spectrum in phase 1. The left
  panel shows the fit with an exponentially cutoff power law plus a
  soft blackbody and high photo-electric absorption. The power law and
  black body components are individually shown. The right panel shows
  an alternative solution consisting of a single power law (the cutoff
  moves beyond the upper energy cutoff of the data) plus highly
  absorbed, partially covered power law with a partial covering
  fraction of 50$\%$. The overall, ISM contribution to the
  photo-electric absorption is low.  The power law and covered power
  law are individually shown.}
\label{cont_01}
\end{figure}

We fit spectra for the five phase bins described in Sect. 3.2 and find
that it can be fit well at most phases using a cut-off power law plus
a soft black body component. The norm variations in both components
follow the flux change in the light curve. The power law index is very
flat with values in the range $\approx -0.4$--$-0.2$.  This flat power
law reflects the fact that we likely observe indirect emission
affected by a central hot corona as already found in \xte and \asca
data~\citep{heinz2001}. The blackbody component has an average
temperature of 181$\pm$11 eV with slight variations, which possibly
reflects some correlation with the photo-electric absorption function
in the fit.  We detect some significant absorption using the updated
\emph{Tbnew} function in \emph{Xspec} of (1.6$\pm0.3)\times10^{21}$
cm$^{-2}$ throughout all phases, which is near the upper limit
presented by ~\citet{heinz2001}. The highest blackbody flux in the
light curve amounts to 3$\times10^{34}$ \ergsec, which we estimate is
only about 2$\%$ of the total flux (see below) and provides an
emission radius of about 20 km.  Figure~\ref{cont_01} (left) shows the
continuum fit to phase 1 with blackbody, power-law, and gaussian line
components.

Alternatively it has also been suggested by ~\citet{heinz2001} and
others that the spectra should be fit using the power law with
additional partial covering absorption.  We have simultaneously fit
phases 1--5 with such a function wherein we forced the power law slope
to be the same for all phases (the fit yields $\Gamma=0.656\pm0.002$,
with 90\% confidence error bars), but let the power law normalization,
line parameters, partial covering fraction, and covering column be
free parameters for each phase.  Figure~\ref{cont_01} (right) shows
the results of this fit for the phase 1 spectra, with the
photo-ionized lines represented by gaussian functions. We find that
this model simultaneously fits all 5 spectra very well
($\chi^2=2266.6/2176$ DoF). The direct and covered power law represent
a normalized flux that ranges from 0.57--1.  That is, at least 43\% of
the total power law flux is completely blocked in the faintest phase,
i.e., phase 4.  (In all likelihood, a larger fraction of the ADC flux
is completely blocked, as phase 1 is unlikely to be completely
uncovered by the disk rim.)  The partial covering fraction varies
between 49--52\%, while the partial covering column varies between
5.4--$6.3\times10^{22}\,{\rm cm}^{-2}$.  In this case the ISM
absorption remains very low at (2.3$\pm0.3)\times10^{20}$ cm$^{-2}$.

The line emitting region is best viewed during phases 1 and 5, with
the least dynamical smear during phase 5. During phase 1 the line
emitting region is more directly viewed, but here the lines switch
from red-shift to blue-shift and the region is likely already partially
obscured by the central corona. Figure~\ref{fig-tri_NeIX} thus shows
the \neix\ line strongest and most narrow. We select phase bin 5 for
the more detailed spectral modeling. The main goal of this analysis is
to get an estimate for the ionization parameter of the emitting
region. Photo-ionization properties are generally described by the
parameter $\xi = L/(n*r^2)$, where L is the source X-ray luminosity, n
is the electron density, and r is the distance to the illuminating
X-ray source~\citep{kallman1982}. From Figure~\ref{cont_01} we already
see that the \ovii\ and \oviii\ line strengths are similar, which is
difficult to achieve with a single ionization parameter. We thus
expect a range of parameters.

We want to keep this analysis as simple as possible and use the new
embedded \emph{Xstar} function \emph{photemis} in
\emph{Xspec}. Fitting the lines with a single ionization parameter, we
obtain log $\xi$ = 2.6 [erg cm/s], which fits the H-like ion lines
very well but not the He-like ion species. However, even though a
second ionization parameter of log $\xi$ = 1.9 [erg cm/s] would fit
these lines, the fit itself is not acceptable as it would produce
significant radiative recombination continua (RRC) for \oviii, \neix~
and \nex, which we do not observe. A viable solution to make up for
this deficiency is to assume a shallow tail of lower ionization
parameters (i.e. below log $\xi$ = 2.6 [erg cm/s]), allowing for high
enough plasma temperatures to smear out the RRCs but produce enough
line flux to fit the He-like lines. The range of this tail depends on
abundance choices and here we cannot find unique solutions.

\section{Discussion}
4U 1822$-$371 is known as one of the prototypical ADC systems and
shows a strong orbital dependence of its X-ray properties.  Our
coverage of many orbital periods allows us to study these dependencies
in detail.  The updated eclipse times from the three \emph{Chandra}
observations are not inconsistent with the trend provided by
previously published X-ray eclipse times~\citep{parmar2000}.  But even
though our result does follow the quadratic function implied by these
previous observations, it does suggest a flatter trend. This is still
consistent with a previous result, which provided the last updated
improved ephemeris of 4U 1822$-$371 derived from UV/optical light
curves~\citep{bayless2009}. But in their analysis, the quadratic term
in the optical is also consistent with the one in X-ray by
\citet{parmar2000} within the measurement errors, and the eclipse time
in the optical lags behind by about 100 seconds. More
recently,~\citet{burderi2010} revisited X-ray observations over the
last 30 yrs confirming the quadratic function by \citet{parmar2000},
but did not include recent optical and UV data. In their analysis the
eclipse times from our \emph{Chandra} observations showed much larger
delays which appear closer to the quadratic function by
\citet{parmar2000}, however the delays also remained significantly
below the function. One possible explanation for the discrepancy with
our result is that they opted not to fold the light curves, which we
think is warranted given the multiple eclipses within these data, and
the fact that we see intermittent dips in the spectra, some occuring
quite close to the main eclipses. In any case, the trend solidifies,
that the orbital period change in this system is larger than expected
from simple magnetic braking and losses due to gravitational
radiation.

The line variations with orbital phase can probe geometrical
properties such as accretion disk size, ADC size, and also a
phase-resolved ionization balance of the line emitting region. While
most of the lower Z ions show significant dynamical properties along
the binary orbit with respect to all of their line properties, the
iron lines only show flux changes consistent with the changes observed
in the light curves.  The Fe K$\alpha$ fluorescence line is visible at
all phases. Its flux is strongest opposite of the eclipse, when the
disk is fully exposed to the observer. During eclipse the fact that
the line flux is still about 30$\%$ of its maximum indicates that the
line emitting region extends significantly beyond the size of the
companion. We also do not believe that the illuminated surface of the
stellar companion is a major source of Fe fluorescence, because at an
edge-on view and opposite to eclipse the disk would likely block the
flux and we would expect maxima during the decreasing and bottom
phases (Table~\ref{tab-fe}). Similarly, using fluorescence
probabilities calculated by ~\citet{bai1979} we also expect maxima at
these phases and not at the opposite (upper) phase. Instead we argue,
consistent with the suggestions by \citet{cottam2001}, that the
fluorescence materials come from an extended region above the disk. In
fact, as~\citet{torrejon2010} showed in a large survey of Fe K
fluorescence in X-ray binaries including 4U 1822$-$371 that the
emissions come from a large, spherically distributed volume around the
accretion disk.  The width of Fe K$\alpha$ fluorescence line are also
better constrained than the previous detection by \citet{cottam2001}
and we determine a lower limit of the emission radius of $R \geq 1
\times 10^{9.5} ~\rm cm$. We do not know how well this fits into the
previous conclusion by~\citet{bayless2009} with respect to the
existence of a layer of cooler material at the base of the wind seen
in the UV, but the X-ray data do not exclude this possibility.

We also detect weak line emissions from hot \fexxvi~ions. As is
observed for the cool Fe K$\alpha$ fluorescence, the \fexxvi, Lyman
$\alpha$ line does not show any line shifts and broadening with
respect to orbital phases.  Its flux roughly follows that of the light
curve and is below detection during eclipse. This indicates that the
line originates from a central region within the disk radius, likely
from parts of the central corona but not the disk itself.  Its
emission volume is then limited by the size of the companion during
eclipse.  Its Roche-lobe filling radius is about 0.54 \Rsun
~\citep{bayless2009} and thus the emission size is less
than 3.5$\times10^{10}$ cm and thus about 0.65 r$_{disk}$ assuming an
orbital separation of a = 1.33$\times10^{11}$ cm and a disk radius 0.4
a (values taken from \citealt{bayless2009}).

Photo-ionized X-ray line emission in 4U 1822$-$371 has been observed
previously~\citep{cottam2001} and phase-resolved spectra suggested
that the emissions from line recombination originate in an X-ray
illuminated bulge located at or near the predicted point of impact of the
accretion stream from the companion and the disk. Our observations
over several binary orbits not only confirm this prediction, but allow
us to study this phenomenon in detail. The lower Z line emissions clearly
show red- and blue-shifted motion with orbital phase which can only be
interpreted by motion of a localized line emitting area at the hot
spot. The maximum absolute velocities from the line shifts of the most
prominent lines (except \fexxvi, see above), i.e., \ovii, \oviii,
\neix, \nex, \mgxi, \mgxii, \sixiv, yield a velocity of 500 \kms,
which is consistent with the orbital speed projection expected from
the orbital parameters of 4U 1822$-$371 at the inclination of
83$^{\circ}$.  Using 40 overlapping phase bins we observe the lines
moving at all times and variations in the line widths appear very
small, which restricts the size of the emitting area to $< 10^6$ cm.
In fact, the very similar width in all lines places the emitting
region at a consistent distance of $1.1\times10^{11}$\,cm to the
source, which is about twice the actual outer disk radius and more
close to the separation of the two stars. However, even though the
uncertainties in the width include the outer disk radius we predict
that there is local turbulence involved in the width.

There are a few more interesting implications from the broadband
analysis of the X-ray spectrum. The fact that we fit a soft blackbody
spectrum to the soft part of the X-ray spectrum is consistent with one
solution obtained from the \asca analysis~\citep{heinz2001}; however,
the blackbody temperature we measure is very much lower. The
properties we determine provide an emission radius close to 20 km,
which appears like the emission from the neutron star itself. However,
the line emission properties imply a radically different picture. As
\citet{heinz2001} already suspected in their analysis of the \asca
data, we likely do not observe the X-ray source directly and the true
source flux should be considerably higher. The ionization parameter of
the line emitting region is directly related to the illuminating
source flux (see Sect.~\ref{photo}). At values between log $\xi$ =
1.9$-$2.6, a disk radius of 5$\times10^{10}$ cm, and a plasma density
between 10$^{12}$ and 10$^{14}$ cm$^{-3}$, for the thin illuminated
layer the illuminating luminosity has to be much larger than $10^{37}$
\ergsec. This also has a consequence for the blackbody emitting
radius, which once adjusted to the higher source luminosity is more of
the order of $>$ 100 km. This makes the solution with the blackbody
component highly unattractive as it removes the possibility that we
see the neutron star, but it is also rather weak to account for inner
disk emissions. In this respect we argue for the solution with the
partial covering component, i.e., the X-ray source is a power law in
nature, likely scattered into our line of sight over the spatially
very extended corona, and due to the edge-on view we observe some of
this emission both blocked and heavily absorbed by the disk rim.

\section{Conclusions}

The orbital phase-resolved analysis of the X-ray photo-ionized region
in 4U~1822$-$37 obtained the following results:

\begin{itemize}
\item{The orbital X-ray ephemeris update suggests a flatter trend with
respect to the previously proposed quadratic change function.}
\item{\fexxvi~ line emission arises from a central corona of size $<3.5\times10^{10}$ cm.}  
\item{Lower Z line emissions show orbital shifts consistent with a small local region
at the rim of the disk, exhibiting orbital blue- and 
red-shifts of the order of 500 \kms.}
\item{The line emitting region has to be the inner illuminated side of a bulge at
the hot spot because it becomes entirely eclipsed when the outer side of the
spot faces the observer.}
\item{The latter fact also implies that the illuminated layer on top of the disk is
thin, at least thin enough not to exceed the outer disk rim.} 
\item{Line widths appear larger than the outer disk radius and likely involve 
local turbulence.}
\item{The line emitting region exhibits high ionization parameters, while missing RRCs
constrain values to be above log $\xi >$ 2. This strongly suggests that we do not directly
observe the X-ray source.}
\item{The luminosity implied by the range of ionization parameters favors an emission
model which consists of a cut-off power law plus a 50$\%$ partial covering absorption.}
\end{itemize}


\acknowledgements
We thank Mike Noble, John Houck for their help with ISIS.  We gratefully acknowledge the 
financial support of {\it Chandra X-Ray Observatory} archive grant AR0-11005X.


\begin{deluxetable}{l|cccl}
\scriptsize
\tablecolumns{5}
\tablecaption{Observations for 4U1822$-$371}
\tablewidth{0cm}
\tablehead{
obsID & MJD interval  & Obs. Start & Exp.(ks) & PI }
\startdata
671  &    51779.681 - 51780.163 & 2000-08-23 16:20:37 UT & 39.95 & Kahn, M.\\
9076 &    54606.949 - 54607.708 & 2008-05-20 22:46:21 UT & 63.68 & Canizares, C.\\
9858 &    54609.551 - 54610.524 & 2008-05-23 13:14:05 UT & 82.19 & Canizares, C.\\
\enddata
\label{tab-obsid}
\end{deluxetable}

\begin{deluxetable}{l|ccc}
\scriptsize
\tablecolumns{4}
\tablecaption{New X-ray eclipse times for 4U1822$-$371 with uncertainties given 
at 90\% confidence.}
\tablewidth{0cm}
\tablehead{
obsID & JD$_{\odot}$  & Uncertainty & Cycle  }
\startdata
671  & 2451780.36424 & 0.00079 & 26561 \\   
9076 & 2454607.69102 & 0.00045 & 38742 \\   
9858 & 2454610.24377 & 0.00037 & 38753 \\   
\enddata
\label{tab-eclipse}
\end{deluxetable}

\clearpage


\begin{deluxetable}{l|ccc|ccc|ccc|ccc}
\tabletypesize{\tiny}
\tablecaption{Ly$\alpha $ lines along the phases for 4U1822$-$371\tablenotemark{a}}
\tablehead{
phase & \multicolumn{3}{c}{SiXIV Ly$\alpha$} & \multicolumn{3}{c}{MgXII Ly$\alpha$} & \multicolumn{3}{c}{NeX Ly$\alpha$} & \multicolumn{3}{c}{OVIII Ly$\alpha$} \\
      & $\lambda$ & Flux & $\sigma$ & $\lambda$ & Flux & $\sigma$ &$\lambda$ & Flux & $\sigma$ & $\lambda$ & Flux & $\sigma$ 
}
\startdata
1 0.0	&   $6.185^{+0.005}_{-0.005} $ & $2.6^{+1.5}_{-1.4} $ & 1.0 & 
            $8.435^{+0.016}_{-0.013} $ & $2.7^{+2.3}_{-1.8} $ & $14.0^{+9.0}_{-6.3} $\tablenotemark{b}& 
            --- & --- & --- &
            $18.997^{+0.037}_{-0.035}$ & $7.5^{+1.3}_{-6.6} $ & $13.2^{+27.7}_{-11.5} $\tablenotemark{b}\\

2 0.025	&   $6.189^{+0.008}_{-0.005} $ & $2.4^{+1.5}_{-1.2} $ & $<15.0 $ &
	    $8.441^{+0.005}_{-0.010} $ & $1.9^{+1.6}_{-1.0} $ & $<13.2 $ &
       	    $12.149^{+0.010}_{-0.005}$ & $2.9^{+2.6}_{-1.4} $ & $<19 $ & 	
	    $18.994^{+0.007}_{-0.008}$ & $9.6^{+6.9}_{-5.2} $ & $<16.4 $ \\

3 0.05	&   $6.190^{+0.005}_{-0.005} $ & $2.7^{+1.1}_{-1.0} $ & 1.0 & 
	    $8.436^{+0.010}_{-0.010}  $ &  $2.7^{+1.6}_{-1.4} $ & $13.2^{+11.2}_{-9.3} $ &
	    $12.149^{+0.010}_{-0.010} $ &  $5.2^{+2.6}_{-2.6} $ & $16.6^{+6.7}_{-6.0} $\tablenotemark{b}&
	    $18.995^{+0.012}_{-0.008}$ & $12.8^{+9.2}_{-7.3} $ & $<18.3 $ \\

4 0.075	&  $6.191^{+0.004}_{-0.003} $ & $4.0^{+1.6}_{-1.4} $ & $4.8^{+3.1}_{-3.1} $\tablenotemark{b} & 
	   $8.433^{+0.009}_{-0.009}  $ &  $2.9^{+1.7}_{-1.4} $ & $13.3^{+11.8}_{-9.9} $ &
	   $12.144^{+0.009}_{-0.009} $ &  $7.1^{+2.8}_{-2.5} $ & $19.1^{+9.8}_{-8.6} $ &
	   $18.993^{+0.007}_{-0.009}$ & $6.8^{+5.9}_{-4.3} $ & 1.0  \\

5 0.1	&  $6.191^{+0.003}_{-0.003} $ & $4.1^{+1.6}_{-1.4} $ & $<9.4 $ & 
	   $8.430^{+0.007}_{-0.007}  $ &  $3.2^{+1.6}_{-1.4} $ & $10.8^{+8.4}_{-8.6} $ &
	   $12.145^{+0.009}_{-0.008} $ &  $8.6^{+3.1}_{-2.7} $ & $20.1^{+9.7}_{-7.0} $ &
	   $18.993^{+0.011}_{-0.009}$ & $12.3^{+8.8}_{-6.1} $ & $<26.1 $ \\

6 0.125	&  $6.192^{+0.003}_{-0.003} $ & $3.8^{+1.5}_{-1.3} $ & $<7.8 $ & 
	   $8.432^{+0.006}_{-0.005}  $ &  $3.0^{+1.5}_{-1.3} $ & $7.9^{+4.5}_{-4.3} $\tablenotemark{b}&
	   $12.145^{+0.007}_{-0.007} $ &  $7.5^{+2.8}_{-2.6} $ & $15.4^{+7.9}_{-7.4} $ &
	   $18.999^{+0.009}_{-0.008}$ & $16.4^{+9.0}_{-7.1} $ & $10.4^{+7.3}_{-6.0} $\tablenotemark{b} \\

7 0.15	&  $6.190^{+0.004}_{-0.004} $ & $4.1^{+1.8}_{-1.5} $ & $4.9^{+3.6}_{-4.9} $\tablenotemark{b} & 
	   $8.430^{+0.006}_{-0.006}  $ &  $3.7^{+1.9}_{-1.6} $ & $8.9^{+5.9}_{-6.3} $\tablenotemark{b}&
           $12.144^{+0.008}_{-0.007} $ &  $7.3^{+2.9}_{-2.7} $ & $15.3^{+8.6}_{-10.5} $ &
	   $18.992^{+0.011}_{-0.010}$ & $22.1^{+10.5}_{-8.8} $ & $18.0^{+14.6}_{-11.2} $\\

8 0.175	&  $6.187^{+0.005}_{-0.005} $ & $4.5^{+2.0}_{-1.8} $ & $8.6^{+6.2}_{-6.2} $ & 
	   $8.430^{+0.004}_{-0.004}  $ &  $3.1^{+2.0}_{-1.1} $ & $<8.7 $&
	   $12.146^{+0.007}_{-0.007} $ &  $8.0^{+2.8}_{-2.5} $ & $16.2^{+6.8}_{-5.0} $ &
	   $18.994^{+0.013}_{-0.013}$ & $26.0^{+12.1}_{-9.9} $ & $25.0^{+20.5}_{-12.5} $\\

9 0.2	&  $6.183^{+0.006}_{-0.006} $ & $4.2^{+2.2}_{-1.8} $ & $9.3^{+9.2}_{-6.2} $ & 
	   $8.427^{+0.005}_{-0.006}  $ &  $4.1^{+1.8}_{-1.8} $ & $9.5^{+4.2}_{-6.4} $\tablenotemark{b}&
	   $12.143^{+0.006}_{-0.006} $ &  $9.2^{+2.8}_{-2.6} $ & $16.1^{+6.0}_{-4.8} $ &
	   $18.983^{+0.020}_{-0.015}$ & $32.4^{+13.9}_{-11.4} $ & $40.1^{+20.7}_{-15.1} $\\

10 0.225&  $6.181^{+0.005}_{-0.005} $ & $5.3^{+2.0}_{-1.8} $ & $9.7^{+5.5}_{-4.0} $ & 
	   $8.428^{+0.004}_{-0.004}  $ &  $3.1^{+1.5}_{-1.2} $ & $<8.9 $&
	   $12.140^{+0.006}_{-0.006} $ &  $9.7^{+2.9}_{-2.6} $ & $16.3^{+5.6}_{-4.9} $ &
	   $18.983^{+0.015}_{-0.015}$ & $22.9^{+11.6}_{-9.9} $ & $28.2^{+22.2}_{-16.2} $\\

11 0.25	&  $6.179^{+0.004}_{-0.004} $ & $5.3^{+2.0}_{-1.8} $ & $8.9^{+4.8}_{-3.7} $ & 
	   $8.428^{+0.007}_{-0.008}  $ &  $3.8^{+1.8}_{-1.6} $ & $12.8^{+8.4}_{-7.7} $&
	   $12.146^{+0.008}_{-0.008} $ &  $7.8^{+2.8}_{-2.6} $ & $17.1^{+7.2}_{-5.7} $ &
	   $18.980^{+0.019}_{-0.017}$ & $21.4^{+11.2}_{-9.3} $ & $30.1^{+19.7}_{-13.4} $\\

12 0.275&  $6.178^{+0.006}_{-0.005} $ & $4.5^{+2.0}_{-1.8} $ & $9.1^{+6.1}_{-4.9} $ & 
	   $8.427^{+0.007}_{-0.008}  $ &  $4.7^{+1.9}_{-1.7} $ & $15.3^{+7.9}_{-6.5} $&
           $12.137^{+0.009}_{-0.009} $ &  $7.9^{+3.0}_{-2.7} $ & $19.5^{+7.8}_{-6.2} $ &
	   $18.974^{+0.023}_{-0.022}$ & $24.5^{+12.5}_{-10.8} $ & $40.3^{+23.1}_{-19.9} $\\

13 0.3	&  $6.176^{+0.006}_{-0.006} $ & $4.7^{+2.2}_{-1.8} $ & $9.9^{+7.3}_{-7.4} $ & 
	   $8.428^{+0.007}_{-0.007}  $ &  $4.3^{+1.9}_{-1.7} $ & $13.8^{+8.3}_{-7.0} $&
	   $12.136^{+0.009}_{-0.008} $ &  $8.3^{+3.0}_{-2.7} $ & $18.6^{+8.3}_{-6.6} $ &
	   $18.957^{+0.013}_{-0.013}$ & $18.4^{+8.9}_{-7.8} $ & $19.3^{+14.8}_{-8.5} $\\

14 0.325&  $6.175^{+0.008}_{-0.007} $ & $3.4^{+1.7}_{-1.9} $ & $8.9^{+4.3}_{-4.4} $\tablenotemark{b} & 
	   $8.426^{+0.008}_{-0.010}  $ &  $4.1^{+2.3}_{-1.8} $ & $15.3^{+14.2}_{-8.6} $&
	   $12.136^{+0.008}_{-0.008} $ &  $9.4^{+3.2}_{-2.8} $ & $19.1^{+8.2}_{-6.5} $ &
	   $18.945^{+0.011}_{-0.010}$ & $14.7^{+9.4}_{-7.6} $ & $12.5^{+9.0}_{-11.9} $\tablenotemark{b}\\

15 0.35	&  $6.174^{+0.008}_{-0.008} $ & $3.6^{+2.3}_{-2.3} $ & $10.1^{+9.4}_{-9.4} $ & 
	   $8.425^{+0.008}_{-0.008}  $ &  $3.9^{+2.1}_{-1.7} $ & $13.8^{+11.5}_{-8.2} $&
	   $12.138^{+0.010}_{-0.009} $ &  $8.3^{+3.2}_{-2.9} $ & $20.2^{+8.8}_{-7.3} $ &
	   $18.949^{+0.010}_{-0.010}$ & $11.8^{+8.3}_{-6.2} $ & $6.4^{+16.4}_{-6.3} $\\

16 0.375&  $6.174^{+0.010}_{-0.015} $ & $3.0^{+1.1}_{-2.0} $ & $<58.2$ & 
	   $8.421^{+0.010}_{-0.012}  $ &  $4.5^{+2.7}_{-2.0} $ & $17.7^{+17.2}_{-8.4} $&
	   $12.136^{+0.010}_{-0.009} $ &  $8.3^{+3.3}_{-2.9} $ & $18.9^{+9.4}_{-7.3} $ &
	   $18.990^{+0.028}_{-0.044}$ & $11.5^{+14.4}_{-8.9} $ & $25.3^{+18.3}_{-17.0} $\tablenotemark{b}\\

17 0.4	&  $6.174^{+0.004}_{-0.005} $ & $2.5^{+1.3}_{-1.3} $ & 1.0 & 
	   $8.423^{+0.008}_{-0.011}  $ &  $3.8^{+2.3}_{-1.8} $ & $14.7^{+14.1}_{-7.9} $&
	   $12.137^{+0.010}_{-0.009} $ &  $8.7^{+3.4}_{-3.1} $ & $20.0^{+9.8}_{-8.1} $ &
	   $18.970^{+0.019}_{-0.026}$ & $24.5^{+13.8}_{-11.0} $ & $34.8^{+27.4}_{-17.3} $\\

18 0.425&  $6.175^{+0.003}_{-0.004} $ & $3.0^{+1.3}_{-1.3} $ & 1.0 & 
	   $8.420^{+0.011}_{-0.009}  $ &  $2.7^{+1.8}_{-1.4} $ & $11.5^{+13.1}_{-5.7} $&
	   $12.139^{+0.007}_{-0.009} $ &  $7.8^{+3.4}_{-3.2} $ & $15.2^{+12.1}_{-9.9} $ &
	   $18.975^{+0.020}_{-0.024}$ & $19.6^{+11.5}_{-9.6} $ & $31.8^{+24.6}_{-15.5} $\\

19 0.45	&  $6.175^{+0.005}_{-0.004} $ & $3.0^{+1.3}_{-1.3} $ & 1.0 & 
	   $8.414^{+0.011}_{-0.011}  $ &  $2.8^{+1.8}_{-1.5} $ & $10.6^{+6.0}_{-4.5} $\tablenotemark{b}&
	   $12.141^{+0.008}_{-0.009} $ &  $7.8^{+3.3}_{-2.9} $ & $16.2^{+9.7}_{-9.9} $ &
	   $18.981^{+0.016}_{-0.017}$ & $20.8^{+11.0}_{-9.2} $ & $28.5^{+18.7}_{-11.6} $\\

20 0.475&  $6.178^{+0.011}_{-0.007} $ & $3.7^{+77.4}_{-2.1} $ & $9.3^{+7.2}_{-5.5} $ \tablenotemark{b}& 
	   $8.404^{+0.010}_{-0.004}  $ &  $2.2^{+1.1}_{-1.0} $ & $<8.6 $&
	   $12.138^{+0.010}_{-0.010} $ &  $7.1^{+3.2}_{-2.8} $ & $17.9^{+10.4}_{-8.9} $ &
	   $18.987^{+0.017}_{-0.017}$ & $15.3^{+9.6}_{-8.0} $ & $24.0^{+18.2}_{-16.8} $\\

21 0.5	&  $6.175^{+0.003}_{-0.004} $ & $2.8^{+1.3}_{-1.3} $ & 1.0 & 
	   $8.404^{+0.004}_{-0.004}  $ & $1.8^{+l.0}_{-1.0} $& 1.0   &
	   $12.135^{+0.012}_{-0.013} $ &  $6.8^{+3.3}_{-2.9} $ & $20.3^{+12.2}_{-10.0} $ &
	   $18.986^{+0.017}_{-0.017}$ & $16.2^{+9.8}_{-8.0} $ & $25.8^{+17.9}_{-15.4} $\\

22 0.525&  $6.178^{+0.009}_{-0.008} $ & $3.1^{+2.0}_{-1.8} $ & $8.2^{+4.9}_{-4.1} $ \tablenotemark{b}& 
	   $8.401^{+0.004}_{-0.005}  $ & $1.5^{+l.0}_{-0.9} $& 1.0  &
	   $12.131^{+0.019}_{-0.018} $ &  $5.4^{+3.3}_{-2.8} $ & $22.7^{+17.8}_{-11.3} $ &
	   $18.996^{+0.013}_{-0.028}$ & $11.7^{+10.5}_{-6.8} $ & $15.4^{+33.7}_{-15.3} $\tablenotemark{b}\\

23 0.55	&  $6.181^{+0.010}_{-0.009} $ & $3.8^{+2.4}_{-2.0} $ & $13.0^{+5.4}_{-4.3} $ \tablenotemark{b}& 
	   $8.403^{+0.004}_{-0.005}  $  & $1.6^{+l.0}_{-0.9} $& 1.0  &
	   $12.142^{+0.015}_{-0.020} $ &  $4.0^{+3.2}_{-2.4} $ & $16.8^{+23.6}_{-12.5} $ &
	   $18.992^{+0.020}_{-0.020}$ & $20.7^{+11.1}_{-9.6} $ & $34.9^{+20.2}_{-23.7} $\\

24 0.575&  $6.186^{+0.011}_{-0.011} $ & $3.7^{+2.6}_{-2.1} $ & $14.6^{+13.9}_{-10.2} $& 
	   $8.405^{+0.005}_{-0.007}  $  & $1.6^{+l.0}_{-0.9} $ & 1.0 & 
	   $12.139^{+0.020}_{-0.019} $ &  $4.3^{+3.3}_{-2.7} $ & $19.7^{+24.0}_{-10.3} $ &
	   $18.990^{+0.024}_{-0.021}$ & $21.2^{+11.6}_{-10.3} $ & $36.3^{+27.6}_{-24.1} $\\

25 0.6	&  ---& ---& ---& 
           ---&--- &--- &
	   $12.127^{+0.019}_{-0.021} $ &  $1.5^{+1.8}_{-1.3} $ & $16.6^{+12.0}_{-11.4} $ &
	   $18.975^{+0.022}_{-0.022}$ & $16.6^{+9.8}_{-8.4} $ & $33.8^{+22.6}_{-17.8} $ \\

26 0.625&  $6.182^{+0.006}_{-0.007} $ & $1.6^{+1.1}_{-1.1} $ & 1.0& 
	   ---&--- &--- & 
	   $12.129^{+0.013}_{-0.012} $ &  $3.6^{+2.4}_{-2.3} $ & $13.5^{+6.5}_{-10.0} $\tablenotemark{b} &
	   $18.982^{+0.038}_{-0.026}$ & $14.1^{+11.6}_{-7.4} $ & $36.8^{+40.7}_{-23.7} $\\

27 0.65	&  ---& ---& ---& 
	   ---&--- &---  & 
           $12.133^{+0.014}_{-0.016} $ &  $3.8^{+2.4}_{-2.4} $ & $16.6^{+12.4}_{-8.3} $ &
	   $18.976^{+0.022}_{-0.021}$ & $15.6^{+9.8}_{-8.0} $ & $32.1^{+21.1}_{-16.0} $\\

28 0.675&  $6.192^{+0.009}_{-0.010} $ & $3.6^{+2.2}_{-1.8} $ & $13.9^{+10.3}_{-6.9} $& 
	   ---&--- &--- & 
	   $12.127^{+0.019}_{-0.007} $ &  $3.1^{+3.3}_{-1.7} $ & $< 21.6 $ &
	   $18.985^{+0.027}_{-0.027}$ & $13.0^{+9.4}_{-7.7} $ & $34.7^{+24.2}_{-19.0} $\\

29 0.7	&  $6.191^{+0.007}_{-0.008} $ & $3.9^{+2.1}_{-1.7} $ & $12.2^{+8.1}_{-5.9} $& 
	   ---&--- &--- &
	   $12.125^{+0.015}_{-0.006} $ &  $3.6^{+2.8}_{-1.8} $ & $< 15.7 $ &
	   ---&--- &---\\

30 0.725&  $6.191^{+0.007}_{-0.007} $ & $3.0^{+1.8}_{-1.5} $ & $9.0^{+8.8}_{-6.6} $& 
	   ---&--- &--- &
	   $12.124^{+0.008}_{-0.005} $ &  $3.5^{+2.0}_{-1.6} $ & $< 10.9 $ &
	   ---&--- &---  \\

31 0.75	&  $6.190^{+0.010}_{-0.010} $ & $2.7^{+1.9}_{-1.7} $ & $9.8^{+5.9}_{-4.8} $\tablenotemark{b} & 
           ---&--- &--- &
	   $12.122^{+0.010}_{-0.006} $ &  $2.9^{+1.7}_{-1.5} $ & $< 9.0 $ &
	   ---&--- &--- \\

32 0.775&  $6.194^{+0.007}_{-0.008} $ & $2.8^{+1.7}_{-1.6} $ & $7.8^{+4.9}_{-5.8} $\tablenotemark{b} & 
           --- &--- &--- &
	   --- &--- &--- &
	   $19.004^{+0.010}_{-0.009}$ & $3.9^{+4.9}_{-3.1} $ & 1.0\\
        
33 0.8	&  $6.194^{+0.007}_{-0.008} $ & $2.8^{+1.9}_{-1.7} $ & $8.6^{+5.7}_{-6.4} $\tablenotemark{b} & 
           --- &--- &--- &
	   --- &--- &--- &
	   $19.003^{+0.012}_{-0.008}$ & $4.4^{+5.1}_{-3.3} $ & 1.0\\

34 0.825&  $6.190^{+0.010}_{-0.011} $ & $2.3^{+2.1}_{-1.5} $ & $9.7^{+6.8}_{-7.2} $\tablenotemark{b} & 
	   $8.436^{+0.005}_{-0.005} $ & $1.3^{+0.9}_{-0.8} $ & 1.0 &
	   ---  &--- &--- &
	   $19.005^{+0.010}_{-0.009}$ & $4.9^{+5.3}_{-3.6} $ & 1.0\\
          
35 0.85	&  $6.190^{+0.011}_{-0.011} $ & $3.0^{+2.3}_{-1.9} $ & $13.1^{+7.2}_{-6.5} $\tablenotemark{b} & 
           $8.436^{+0.004}_{-0.004} $ & $1.6^{+0.9}_{-0.8} $& 1.0& 
           $12.152^{+0.007}_{-0.007} $ &  $2.0^{+1.5}_{-1.2} $ & 1.0 &
	   $19.003^{+0.010}_{-0.009}$ & $5.2^{+5.6}_{-3.8} $ & 1.0\\

36 0.875&  ---& ---& ---& 
   	   $8.437^{+0.003}_{-0.003} $ & $1.7^{+0.9}_{-0.8} $ & 1.0& 
	   $12.153^{+0.006}_{-0.008} $ &  $2.2^{+1.5}_{-1.2} $ & 1.0 &
	   $19.007^{+0.014}_{-0.023}$ & $8.2^{+7.4}_{-5.7} $ & $<36.0 $\\

37 0.9	&  ---& ---& ---& 
           $8.439^{+0.005}_{-0.004} $  &$1.7^{+0.9}_{-0.8} $  &1.0 &  
           $12.149^{+0.008}_{-0.008} $ &  $2.4^{+2.0}_{-1.3} $ & 1.0 &
	   $19.000^{+0.014}_{-0.026}$ & $5.1^{+6.0}_{-4.1} $ & 1.0\\

38 0.925&  $6.185^{+0.008}_{-0.006} $ & $1.5^{+1.1}_{-1.1} $ & 1.0& 
           $8.439^{+0.005}_{-0.004} $  &$1.9^{+0.9}_{-0.8} $  &1.0&
           ---&--- &--- &
	   $19.001^{+0.010}_{-0.011}$ & $8.1^{+6.7}_{-4.9} $ & 1.0\\

39 0.95	&  $6.185^{+0.007}_{-0.005} $ & $1.6^{+1.1}_{-1.0} $ & 1.0& 
	   $8.436^{+0.008}_{-0.008}  $ &  $3.4^{+1.6}_{-1.4} $ & $13.6^{+8.2}_{-6.8} $ &
           ---&--- &--- &
	   $19.000^{+0.010}_{-0.011}$ & $8.1^{+6.7}_{-4.9} $ & 1.0\\

40 0.975&  $6.185^{+0.004}_{-0.004} $ & $2.2^{+1.1}_{-1.1} $ & 1.0& 
	   $8.436^{+0.008}_{-0.008}  $ &  $3.4^{+1.6}_{-1.4} $ & $13.6^{+8.2}_{-6.8} $\tablenotemark{b}&
           ---&--- &--- &
	   $18.996^{+0.021}_{-0.016}$ & $7.5^{+7.6}_{-5.3} $ & $11.6^{+12.5}_{-7.2}$\tablenotemark{b} \\
\enddata

\tablenotetext{a}{
$\lambda$ is in units of \AA, Flux is in units of $10^{-5}\rm photons ~s^{-1} ~cm^{-2} $, and 
$\sigma$ is in units of  $10^{-3}$ \AA.
}
\tablenotetext{a}{ 67\% confidence values}
\label{tab-lya}
\end{deluxetable}

\clearpage



\begin{deluxetable}{l|ccc|ccc|ccc}
\tabletypesize{\footnotesize}
\tablecaption{Intercombination lines along the phases for 4U1822$-$371\tablenotemark{a}}
\tablehead{
phase &  \multicolumn{3}{c}{MgXI i} & \multicolumn{3}{c}{NeIXi} & \multicolumn{3}{c}{OVII i} \\
      &  $\lambda$ & Flux & $\sigma$ &$\lambda$ & Flux & $\sigma$ & $\lambda$ & Flux & $\sigma$ 
}
\startdata
1 0.0	&  $9.243^{+0.006}_{-0.008}  $ &  $1.5^{+1.3}_{-1.1} $  & 1.0 & ---  &---  & ---  & 
	   $21.830^{+0.009}_{-0.009} $ &  $21.3^{+22.1}_{-13.7} $ &  1.0 \\

2 0.025	&  $9.241^{+0.009}_{-0.013}  $ &  $1.6^{+1.4}_{-1.1} $ & $<13.1 $ & 
           $13.549^{+0.016}_{-0.011} $ &  $6.5^{+3.4}_{-3.2} $ & $17.7^{+14.1}_{-9.7} $ &
	   $21.826^{+0.011}_{-0.009} $ &  $11.7^{+12.4}_{-8.3} $ &  1.0 \\

3 0.05	&  $9.243^{+0.005}_{-0.008}  $ &  $2.4^{+1.7}_{-1.1} $ & $<10.1 $ & 
           $13.557^{+0.012}_{-0.011} $ &  $5.8^{+3.6}_{-2.6} $ & $12.9^{+13.6}_{-7.1} $ &
	   $21.835^{+0.006}_{-0.007} $ &  $18.4^{+14.2}_{-10.1} $ &  1.0 \\

4 0.075	&  $9.233^{+0.011}_{-0.010}  $ &  $4.1^{+2.0}_{-1.9} $ & $16.7^{+11.7}_{-6.8} $ & 
	   $13.561^{+0.007}_{-0.008} $ &  $7.1^{+3.3}_{-2.8} $ & $11.4^{+5.1}_{-4.6} $\tablenotemark{b} &
	   $21.836^{+0.009}_{-0.006} $ &  $18.8^{+14.1}_{-10.1} $ &  1.0 \\

5 0.1	&  $9.241^{+0.005}_{-0.016}  $ &  $2.7^{+2.3}_{-1.2} $ & $<10.6 $& 
	   $13.565^{+0.007}_{-0.008} $ &  $8.3^{+3.6}_{-3.1} $ & $12.9^{+8.9}_{-6.9} $ &
	   $21.837^{+0.009}_{-0.007} $ &  $15.4^{+13.5}_{-9.2} $ &  1.0 \\

6 0.125	&  $9.232^{+0.006}_{-0.007}  $ &  $5.6^{+2.0}_{-1.8} $ & $15.4^{+7.0}_{-5.6} $& 
	   $13.565^{+0.005}_{-0.005} $ &  $9.7^{+3.6}_{-3.2} $ & $9.5^{+7.3}_{-6.1} $ &
	   $21.825^{+0.016}_{-0.015} $ &  $37.4^{+20.9}_{-17.3} $ & $25.2^{+8.7}_{-7.9} $\tablenotemark{b} \\

7 0.15	&  $9.234^{+0.009}_{-0.004}  $ &  $5.2^{+2.0}_{-2.2} $ & $12.7^{+6.3}_{-8.8} $ & 
           $13.564^{+0.004}_{-0.005} $ &  $13.8^{+4.4}_{-3.8} $ & $9.9^{+7.1}_{-7.0} $ &
	   $21.824^{+0.015}_{-0.013} $ &  $42.3^{+21.7}_{-17.5} $ & $24.5^{+14.7}_{-11.5} $\\

8 0.175	&  $9.238^{+0.003}_{-0.004}  $ &  $5.0^{+1.7}_{-1.5} $ & $7.7^{+5.2}_{-4.4} $& 
           $13.562^{+0.005}_{-0.005} $ &  $15.0^{+4.3}_{-3.8} $ & $14.0^{+6.2}_{-4.7} $ &   
	   $21.814^{+0.011}_{-0.011} $ &  $61.3^{+24.6}_{-20.7} $ & $26.6^{+12.4}_{-7.7} $\\

9 0.2	&  $9.236^{+0.005}_{-0.004}  $ &  $4.4^{+1.7}_{-1.5} $ & $7.9^{+5.9}_{-7.8} $ & 
	   $13.560^{+0.006}_{-0.006} $ &  $15.5^{+4.7}_{-4.1} $ & $17.4^{+8.2}_{-6.3} $ &
	   $21.808^{+0.010}_{-0.011} $ &  $62.5^{+24.8}_{-20.7} $ & $25.1^{+13.0}_{-7.7} $\\

10 0.225&  $9.236^{+0.005}_{-0.005}  $ &  $4.6^{+1.7}_{-1.6} $ & $8.0^{+3.6}_{-3.7} $\tablenotemark{b} &
 	   $13.557^{+0.009}_{-0.008} $ &  $15.9^{+5.2}_{-4.5} $ & $24.1^{+11.2}_{-7.5} $ &
	   $21.796^{+0.013}_{-0.014} $ &  $63.6^{+26.0}_{-21.9} $ & $29.5^{+16.2}_{-10.4} $\\

11 0.25	&  $9.232^{+0.005}_{-0.007}  $ &  $4.3^{+2.1}_{-1.8} $ & $7.8^{+6.3}_{-5.6} $\tablenotemark{b} &
	   $13.551^{+0.006}_{-0.006} $ &  $18.1^{+5.2}_{-4.6} $ & $18.5^{+9.1}_{-6.9} $ &
	   $21.793^{+0.011}_{-0.012} $ &  $65.3^{+25.9}_{-21.9} $ & $27.4^{+12.9}_{-9.0} $\\

12 0.275& $9.230^{+0.005}_{-0.005}  $ &  $4.3^{+2.0}_{-1.6} $ & $7.8^{+9.1}_{-6.7} $ & 
	   $13.549^{+0.005}_{-0.005} $ &  $17.8^{+4.8}_{-4.1} $ & $15.2^{+7.4}_{-4.8} $ &
	   $21.789^{+0.013}_{-0.016} $ &  $51.2^{+24.7}_{-20.2} $ & $25.2^{+18.0}_{-10.9} $\\

13 0.3	&  $9.228^{+0.005}_{-0.006}  $ &  $5.8^{+2.4}_{-2.0} $ & $12.4^{+10.0}_{-6.2} $ & 
	   $13.544^{+0.006}_{-0.006} $ &  $15.5^{+5.6}_{-4.2} $ & $15.9^{+12.7}_{-6.4} $ &
	   $21.800^{+0.009}_{-0.032} $ &  $49.4^{+28.0}_{-24.4} $ & $30.5^{+24.5}_{-18.1} $\\

14 0.325&  $9.225^{+0.006}_{-0.006}  $ &  $6.9^{+2.5}_{-2.1} $ & $16.1^{+9.0}_{-5.4} $ & 
	   $13.541^{+0.007}_{-0.007} $ &  $13.9^{+4.8}_{-4.2} $ & $18.0^{+9.7}_{-7.4} $ &
	   $21.798^{+0.018}_{-0.020} $ &  $37.7^{+23.6}_{-19.0} $ & $28.0^{+21.4}_{-15.9} $\\

15 0.35	&  $9.222^{+0.006}_{-0.006}  $ &  $7.4^{+2.7}_{-2.2} $ & $15.5^{+9.2}_{-5.9} $ & 
	   $13.537^{+0.007}_{-0.007} $ &  $16.0^{+4.9}_{-4.4} $ & $20.5^{+9.5}_{-6.7} $ &
	   $21.800^{+0.019}_{-0.021} $ &  $37.8^{+23.4}_{-18.5} $ & $30.4^{+22.5}_{-13.6} $\\

16 0.375&  $9.221^{+0.006}_{-0.006}  $ &  $7.9^{+2.8}_{-2.3} $ & $16.9^{+8.6}_{-6.0} $ & 
	   $13.545^{+0.008}_{-0.009} $ &  $18.3^{+6.0}_{-5.1} $ & $26.7^{+13.2}_{-8.5} $ &
	   $21.806^{+0.015}_{-0.015} $ &  $37.3^{+21.8}_{-17.5} $ & $23.6^{+15.3}_{-11.0} $\\

17 0.4	&  $9.223^{+0.007}_{-0.006}  $ &  $6.8^{+2.6}_{-2.2} $ & $15.3^{+8.8}_{-6.0} $& 
	   $13.545^{+0.008}_{-0.009} $ &  $15.8^{+5.3}_{-4.6} $ & $22.9^{+11.9}_{-7.7} $ &
	   $21.813^{+0.015}_{-0.014} $ &  $30.0^{+19.3}_{-15.1} $ & $20.3^{+13.1}_{-10.8} $\\

18 0.425&  $9.224^{+0.008}_{-0.009}  $ &  $7.6^{+3.2}_{-2.6} $ & $21.3^{+11.9}_{-7.7} $ & 
	   $13.535^{+0.011}_{-0.013} $ &  $13.9^{+7.2}_{-4.9} $ & $27.3^{+25.3}_{-10.7} $ &
	   $21.818^{+0.015}_{-0.015} $ &  $26.1^{+18.3}_{-14.0} $ & $18.9^{+14.0}_{-10.7} $\\

19 0.45	&  $9.225^{+0.010}_{-0.010}  $ &  $7.5^{+3.2}_{-2.6} $ & $24.6^{+12.2}_{-8.0} $ & 
	   $13.540^{+0.011}_{-0.010} $ &  $14.7^{+6.0}_{-4.8} $ & $27.4^{+19.8}_{-9.0} $ &
	   $21.816^{+0.013}_{-0.013} $ &  $26.5^{+18.1}_{-14.0} $ & $16.8^{+6.7}_{-6.4} $\tablenotemark{b}\\

20 0.475&  $9.227^{+0.012}_{-0.012}  $ &  $6.3^{+3.4}_{-2.6} $ & $24.6^{+17.2}_{-9.8} $ & 
	   $13.540^{+0.015}_{-0.010} $ &  $18.3^{+6.5}_{-5.4} $ & $33.3^{+15.4}_{-10.1} $ &
	   $21.811^{+0.020}_{-0.021} $ &  $21.1^{+17.8}_{-14.2} $ & $21.5^{+10.4}_{-8.9} $\tablenotemark{b}\\

21 0.5	&  $9.216^{+0.010}_{-0.006}  $ &  $1.8^{+1.2}_{-1.1} $ & 1.0 & 
	   $13.544^{+0.011}_{-0.015} $ &  $16.1^{+10.6}_{-5.1} $ & $29.8^{+33.8}_{-10.7} $ &
	   $21.801^{+0.021}_{-0.051} $ &  $21.8^{+18.2}_{-13.9} $ & $23.3^{+27.6}_{-19.5} $\\

22 0.525&  --- &--- &---& 
	   $13.548^{+0.009}_{-0.009} $ &  $14.3^{+5.1}_{-4.3} $ & $24.3^{+14.8}_{-7.0} $ &
	   $21.814^{+0.008}_{-0.009} $ &  $12.4^{+13.0}_{-8.4} $ & 1.0\\

23 0.55	&  --- &--- &--- & 
	   $13.550^{+0.011}_{-0.012} $ &  $12.1^{+5.0}_{-4.2} $ & $24.9^{+14.7}_{-8.4} $ &
	   $21.819^{+0.007}_{-0.008} $ &  $14.7^{+14.0}_{-9.4} $ & 1.0\\

24 0.575&  --- &--- &--- & 
	   $13.554^{+0.010}_{-0.010} $ &  $11.7^{+4.8}_{-4.1} $ & $21.9^{+13.2}_{-8.5} $ &
	   $21.818^{+0.007}_{-0.007} $ &  $19.4^{+15.7}_{-11.0} $ & 1.0\\

25 0.6	&  --- &--- &--- & 
	   $13.560^{+0.010}_{-0.010} $ &  $9.7^{+4.3}_{-3.7} $ & $19.4^{+13.7}_{-8.6} $ &
	   $21.809^{+0.019}_{-0.020} $ &  $45.9^{+24.3}_{-20.1} $ & $35.8^{+20.4}_{-12.7} $\\

26 0.625&  --- &--- &---& 
	   $13.562^{+0.011}_{-0.012} $ &  $9.8^{+4.7}_{-3.8} $ & $22.4^{+16.6}_{-10.2} $ &
	   $21.810^{+0.019}_{-0.019} $ &  $48.1^{+24.3}_{-20.3} $ & $37.2^{+19.7}_{-12.2} $\\

27 0.65	&  --- &--- &---& 
	   $13.566^{+0.008}_{-0.009} $ &  $9.0^{+4.0}_{-3.4} $ & $16.4^{+11.6}_{-7.8} $ &
	   $21.818^{+0.020}_{-0.023} $ &  $42.7^{+24.0}_{-20.2} $ & $33.9^{+24.6}_{-16.9} $\\

28 0.675&  --- &--- &--- & 
	   $13.565^{+0.010}_{-0.011} $ &  $7.3^{+3.7}_{-3.1} $ & $17.0^{+12.8}_{-9.0} $ &
	   $21.825^{+0.014}_{-0.018} $ &  $39.2^{+22.8}_{-17.9} $ & $23.8^{+19.7}_{-10.9} $\\

29 0.7	&  --- &--- &---& 
	   $13.567^{+0.011}_{-0.014} $ &  $7.4^{+3.6}_{-3.1} $ & $18.4^{+14.2}_{-8.7} $ &
	   $21.822^{+0.015}_{-0.020} $ &  $34.2^{+21.4}_{-16.6} $ & $23.0^{+21.1}_{-11.0} $\\

30 0.725&  $9.223^{+0.007}_{-0.006} $ & $1.2^{+1.0}_{-0.9} $ & 1.0& 
	   $13.562^{+0.014}_{-0.020} $ &  $7.8^{+4.7}_{-3.5} $ & $24.5^{+21.9}_{-11.8} $ &
	   $21.826^{+0.015}_{-0.016} $ &  $29.6^{+18.5}_{-14.2} $ & $23.1^{+18.4}_{-9.8} $\\

31 0.75	&  $9.224^{+0.007}_{-0.006} $ & $1.2^{+1.0}_{-0.9} $ & 1.0& 
           ---& ---& ---& 
	   $21.850^{+0.010}_{-0.009} $ &  $14.7^{+14.0}_{-9.4} $ & 1.0\\

32 0.775&  --- &--- &--- &--- &--- &--- &--- &--- &---\\

33 0.8	&  --- &--- &--- &--- &--- &--- &--- &--- &---\\

34 0.825&  $9.229^{+0.007}_{-0.007} $ &  $1.0^{+0.9}_{-0.8} $  & 1.0 & ---  &--- &--- & --- &--- &---\\
          
35 0.85	&  $9.228^{+0.006}_{-0.004} $ & $1.3^{+0.9}_{-0.8} $  & 1.0 & ---&--- &--- &  --- &--- &---\\

36 0.875&  $9.228^{+0.005}_{-0.004} $ & $1.3^{+0.9}_{-0.8} $ & 1.0& 
           $13.561^{+0.005}_{-0.006} $  & $3.4^{+2.2}_{-1.8} $ & 1.0 & 
	   $21.817^{+0.020}_{-0.020} $ &  $33.1^{+19.4}_{-15.4} $ & $34.7^{+19.5}_{-12.3} $\\

37 0.9	&  $9.229^{+0.006}_{-0.005} $  &$1.1^{+0.9}_{-0.8} $   &1.0 & 
           $13.566^{+0.013}_{-0.011} $ &  $6.3^{+3.7}_{-3.1} $ & $15.8^{+15.2}_{-14.1} $ & 
	   $21.820^{+0.030}_{-0.030} $ &  $22.1^{+19.3}_{-14.8} $ & $30.6^{+15.9}_{-10.7} $\tablenotemark{b}\\

38 0.925&  $9.231^{+0.004}_{-0.005} $  & $1.6^{+1.0}_{-0.9} $   &1.0 & 
	   $13.560^{+0.012}_{-0.011} $ &  $4.7^{+8.3}_{-2.9} $ & $<34.7 $ &
	   $21.822^{+0.024}_{-0.025} $ &  $20.4^{+17.6}_{-13.4} $ & $24.3^{+12.0}_{-8.1} $\tablenotemark{b}\\

39 0.95	&  $9.231^{+0.005}_{-0.006} $  & $1.4^{+0.9}_{-0.8} $  &1.0 & 
	   $13.560^{+0.014}_{-0.051} $ &  $4.5^{+10.2}_{-2.9} $ & $14.4^{+14.4}_{-11.8} $\tablenotemark{b} &
	   $21.831^{+0.016}_{-0.018} $ &  $26.0^{+18.3}_{-13.9} $ & $21.5^{+18.0}_{-10.1} $\\

40 0.975&  $9.235^{+0.009}_{-0.010}  $ &  $1.9^{+1.3}_{-1.2} $ & $9.6^{+5.0}_{-4.3} $\tablenotemark{b} & 
	   $13.560^{+0.014}_{-0.051} $ &  $4.5^{+10.2}_{-2.9} $ & $14.4^{+9.5}_{-9.8} $\tablenotemark{b} &
	   $21.832^{+0.014}_{-0.016} $ &  $24.2^{+17.5}_{-13.2} $ & $17.5^{+16.9}_{-10.6} $\\
	
\enddata

\tablenotetext{a}{
$\lambda$ is in units of \AA, Flux is in units of $10^{-5}\rm photons ~s^{-1} ~cm^{-2} $, and 
$\sigma$ is in units of  $10^{-3}$ \AA.
}
\tablenotetext{a}{ 67\% confidence levels}
\label{tab-I}
\end{deluxetable}

\clearpage

\begin{deluxetable}{l|ccc|ccc}
\tabletypesize{\small}
\tablewidth{0cm}
\tablecaption{Fe  lines along the phases for 4U1822$-$371\tablenotemark{a}}
\tablehead{
phase &  \multicolumn{3}{c}{FeK$\alpha $}  & \multicolumn{3}{c}{FeXXVI Ly$\alpha$} \\
\cline{2-7}
      &  $\lambda$ & Flux & $\sigma$ &$\lambda$ & Flux & $\sigma$ 
}
\startdata
1(upper)&  $1.937^{+0.001}_{-0.001}$ & $46.7^{+10.0}_{-8.0} $  & $5.7^{+2.5}_{-2.2}$ & 
	   $1.782^{+0.003}_{-0.003}$  & $20.8^{+10.0}_{-8.0} $  & $<7.1$ \\
	
2(decreasing)	&  $1.944^{+0.008}_{-0.009}$  & $12.2^{+10.6}_{-10.0} $  & 1.0 &
	   $1.786^{+0.002}_{-0.004}$  & $10.2^{+11.6}_{-10.2} $  &  1.0 \\

3(bottom)  &  $1.938^{+0.003}_{-0.003}$  & $16.8^{+6.4}_{-4.8} $  & $<7.2$ &
	   $1.786^{+0.006}_{-0.005}$  & $7.4^{+4.9}_{-4.9} $  &  1.0 \\
	
4(eclipse)&  $1.936^{+0.007}_{-0.003}$  & $14.9^{+7.5}_{-7.5} $  & 1.0 &
	   --- & ---&--- \\

5(rising)&  $1.937^{+0.002}_{-0.002}$  & $31.6^{+8.3}_{-7.9} $  & $<5.7$ &
	   $1.783^{+0.003}_{-0.004}$  & $19.3^{+7.3}_{-7.3} $  &  1.0 \\

\enddata

\tablenotetext{a}{
$\lambda$ is in units of \AA, Flux is in units of $10^{-5}\rm photons ~s^{-1} ~cm^{-2} $, and
$\sigma$ is in units of  $10^{-3}$ \AA.
}
\label{tab-fe}
\end{deluxetable}

\begin{deluxetable}{l|ccc|cc|c}
\tabletypesize{\small}
\tablewidth{0cm}
\tablecaption{\neix~ triplet line ratios along the phases for 4U1822$-$371}
\tablehead{
phase &  $\lambda$\tablenotemark{a}  & $\sigma$ & flux & G & R & Cash/dof\\
      &  (\AA)      & ($10^{-3}\AA $) & ($10^{-5}\rm photons ~s^{-1} ~cm^{-2} $) & (f+i)/r& f/i& 
}
\startdata
1(upper)&  $13.541^{+0.007}_{-0.005}$ & $24^{+7}_{-5} $  & $15.4^{+3.4}_{-3.2}$ & 
	   $4.8^{+7.8}_{-2.0}$  & $<0.3 $  &  74/81 \\
	
2(decreasing) &  $13.560^{+0.009}_{-0.020}$  & $18^{+16}_{-9} $  & $9.2^{+5.6}_{-4.7}$ &
	  --- & --- &  67/81\\

3(bottom)  &  $13.565^{+0.009}_{-0.009}$  & $20^{+10}_{-7} $  & $6.6^{+2.6}_{-2.1}$ &
	   $>3.2$  & $<0.4$  &  111/81 \\
	
4(eclipse)&  $13.470^{+0.016}_{-0.005}$  & 1.0  & $3.1^{+2.7}_{-2.1}$&
	   --- & ---& 77/81\\

5(rising)&  $13.563^{+0.004}_{-0.004}$  & $14^{+7}_{-5} $  & $14.4^{+3.9}_{-3.3}$ &
	   $4.9^{+10.2}_{-2.3}$  & $<0.2 $  &   65/81\\

\enddata

\tablenotetext{a}{
strongest line in the triplet complex: intercombination line for phase 1,2,3,5; resonance line for phase 4.
}
\label{tab-tri_NeIX}
\end{deluxetable}

\begin{deluxetable}{l|ccc|cc|c}
\tabletypesize{\small}
\tablewidth{0cm}
\tablecaption{Triplet line ratios for two new observations of 4U1822$-$371}
\tablehead{
Ion &  $\lambda(i)$  & $\sigma$ & flux & G & R & Cash/dof\\
      &  (\AA)      & ($10^{-3}\AA $) & ($10^{-5}\rm photons ~s^{-1} ~cm^{-2} $) & (f+i)/r& f/i& 
}
\startdata
\mgxi &  $9.232^{+0.005}_{-0.004}$ & $17^{+5}_{-4} $  & $3.9^{+1.0}_{-0.8}$ & 
	   $>3.8$  & $<0.2 $  &  16/27 \\
	
\neix &  $13.553^{+0.007}_{-0.003}$  & $24^{+5}_{-4} $  & $11.0^{+1.7}_{-1.6}$ &
	 $5.5^{+6.0}_{-2.0}$   & $<0.2$ &  89/81\\

\ovii &  $21.817^{+0.005}_{-0.013}$  & $36^{+10}_{-8} $  & $38.4^{+8.7}_{-7.6}$ &
	   $4.5^{+5.5}_{-1.8}$  & $<0.09 $ &  111/81 \\
	
\enddata

\label{tab-tri_all}
\end{deluxetable}

\clearpage

\begin{figure}
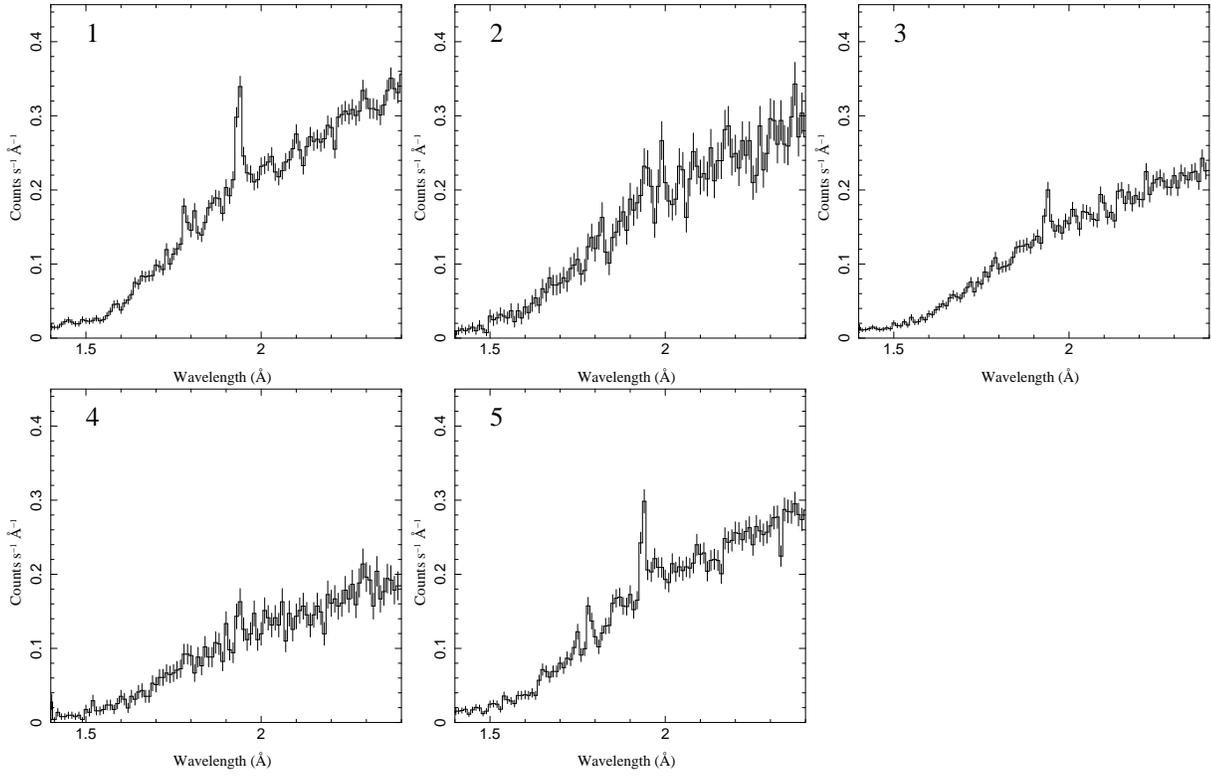

\mbox{
\includegraphics[width=2.0in,angle=270]{newobs_spec_pha1_fe.ps}
\includegraphics[width=2.0in,angle=270]{newobs_spec_pha2_fe.ps}
\includegraphics[width=2.0in,angle=270]{newobs_spec_pha3_fe.ps}
}
\mbox{
\includegraphics[width=2.0in,angle=270]{newobs_spec_pha4_fe.ps}
\includegraphics[width=2.0in,angle=270]{newobs_spec_pha5_fe.ps}
}
\caption{The phase folded spectra for the five phase bins shown in
  Figure~\protect{\ref{fig-ephemeris}}.}
\label{fig-Fe}
\end{figure}


\end{document}